\documentclass{article}

\PassOptionsToPackage{numbers, compress}{natbib}

\usepackage[main,final]{neurips_2026}

\usepackage[utf8]{inputenc} 
\usepackage[T1]{fontenc}    
\usepackage{fontawesome5}   
\usepackage{hyperref}       
\usepackage{url}            
\usepackage{booktabs}       
\usepackage{amsfonts}       
\usepackage{nicefrac}       
\usepackage{microtype}      
\usepackage[table]{xcolor}  
\usepackage{xspace}
\usepackage{tikz}
\usepackage[misc]{ifsym}
\usepackage{amsmath, amssymb, amsthm}
\usepackage{dsfont, pifont}
\usepackage{algorithm, algorithmic}
\usepackage{multirow, multicol, wrapfig}
\usepackage{makecell}
\usepackage{enumitem}
\usepackage{appendix}
\usepackage{setspace}

\definecolor{YaleBlue}{rgb}{0.059,0.302,0.573}
\definecolor{forestgreen}{rgb}{0.133,0.549,0.133}
\definecolor{crimson}{rgb}{0.863,0.078,0.235}
\definecolor{firebrick}{rgb}{0.698,0.133,0.133}
\definecolor{wine}{HTML}{830E0D}

\newcommand{\ours}{\texttt{RNAGenScape}\xspace}
\newcommand{\cmark}{\textcolor{forestgreen!80!black}{\ding{51}}}
\newcommand{\xmark}{\textcolor{crimson!80!black}{\ding{55}}}
\newcommand{\fancynumber}[1]{%
  \tikz[baseline=(char.base)]{
    \node[shape=circle,draw=white,fill=wine!90,inner sep=1pt](char){\color{white}#1};
  }%
}
\definecolor{citeblue}{RGB}{30,80,160}
\hypersetup{
  colorlinks=true,
  linkcolor=citeblue,
  citecolor=citeblue,
  urlcolor=citeblue,
}

\title{\texttt{RNAGenScape}: Property-Guided, Optimized Generation of mRNA Sequences with\\Manifold Langevin Dynamics}

\author{
\textbf{Danqi Liao}$^{*,1}$ \quad
\textbf{Chen Liu}$^{*,1}$ \quad
\textbf{Xingzhi Sun}$^{*,1}$ \quad
\textbf{Di\'e Tang}$^{2}$ \quad
\textbf{Haochen Wang}$^{2}$\\
\textbf{Scott Youlten}$^{2}$ \quad
\textbf{Srikar Krishna Gopinath}$^{2}$ \quad
\textbf{Haejeong Lee}$^{2}$ \quad
\textbf{Ethan C. Strayer}$^{2}$ \\
\textbf{Antonio J. Giraldez}$^{2}$\textsuperscript{~\Letter} \quad
\textbf{Smita Krishnaswamy}$^{1,2,3}$\textsuperscript{~\Letter} \vspace{4pt}\\
$^{1}$Department of Computer Science, Yale University \quad $^{2}$Department of Genetics, Yale University\\
$^{3}$Center for Neurocomputation and Machine Intelligence, Wu Tsai Institute \quad $^{*}$Equal contribution\\
\footnotesize
\faGithub~\href{https://github.com/KrishnaswamyLab/RNAGenScape}{$\texttt{KrishnaswamyLab/RNAGenScape}$}
\hfill
\textsuperscript{\Letter}\url{{antonio.giraldez, smita.krishnaswamy}@yale.edu}
}

\begin{document}

\maketitle

\begin{abstract}
Generating property-optimized mRNA sequences is central to applications such as vaccine design and protein replacement therapy, but remains challenging due to limited data, complex sequence-function relationships, and the narrow space of biologically viable sequences. Generative methods that drift away from the data manifold can yield sequences that fail to fold, translate poorly, or are otherwise nonfunctional. We present \ours, a property-guided manifold Langevin dynamics framework for mRNA sequence generation that operates on a learned manifold of real data. By performing iterative local optimization constrained to this manifold, \ours preserves favorable computational viability proxies, accesses reliable guidance, and discourages excursions into regions poorly supported by the training data. The framework integrates three components: \fancynumber{1}~an autoencoder jointly trained with a property predictor to learn a property-organized latent manifold, \fancynumber{2}~a denoising autoencoder that projects updates back onto the manifold, and \fancynumber{3}~a property-guided Langevin dynamics procedure that performs optimization along the manifold. Across three real-world mRNA datasets spanning two orders of magnitude in size, \ours increases median oracle-predicted property gain by up to 267\% and success rate by up to 31\%, while maintaining good biological viability proxy measures and achieving competitive inference efficiency compared to existing generative approaches.
\end{abstract}

\addtocontents{toc}{\protect\setcounter{tocdepth}{-1}}
\vspace{-4pt}
\section{Introduction}
\vspace{-4pt}

Messenger ribonucleic acids~(mRNAs) have emerged as a core modality for modern therapeutics, enabling applications ranging from mRNA vaccine design and protein replacement therapy to gene regulation and synthetic biology~\cite{pardi2018mrna, chaudhary2021mrna, qin2022mrna, vavilis2023mrna}. In these settings, the nucleotide sequence of an mRNA directly determines properties such as stability, translational efficiency, and ultimately protein yield. Even small sequence edits, particularly in non-coding regions such as the 5' untranslated region~(UTR), can lead to substantial changes in degradation rates and ribosome loading~\cite{castillo2024optimizing, ma2024optimization, li2025optimizing}. As a result, systematic optimization of the mRNA sequence has the potential to improve therapeutic efficacy, reduce dosage requirements, and expand the design space of viable mRNA-based interventions.

Despite its importance, mRNA sequence generation and optimization remains underexplored from a machine learning perspective~\cite{castillo2021machine, schlusser2024current}. A key challenge is that viable mRNA sequences occupy a narrow and structured subset of the astronomically large ambient sequence space~\cite{calvanese2024towards, zhang2023algorithm}. Random or unconstrained edits can lead to nonfunctional transcripts, while experimentally measured datasets are typically small, imbalanced, and sparsely cover the space of interest~\cite{taubert2023rna, asim2025rna}. Moreover, sequence-function relationships in mRNA are highly nonlinear and context-dependent, making the optimization of mRNA sequences a challenging problem~\cite{licatalosi2010resolving, weinreb20163d}.

To address the challenges, we introduce \ours, a property-guided manifold Langevin dynamics framework for mRNA sequence generation built around three key components. \fancynumber{1}~An autoencoder and a property predictor, together termed an organized autoencoder~(OAE), that jointly learn a latent manifold of mRNA sequences while organizing this manifold according to the target property. This latent manifold represents the subspace of viable mRNA sequences rather than the much larger Euclidean ambient space. \fancynumber{2}~A manifold projector, implemented as a denoising autoencoder, that maps data onto the learned manifold and prevents drift toward implausible regions. To ensure robustness under sparse and imbalanced data regimes common in mRNA studies, we use SUGAR~\cite{SUGAR} to fill holes in the data manifold and stabilize manifold projection. \fancynumber{3}~A property-guided manifold Langevin dynamics that iteratively refines sequences by combining gradient guidance with stochastic exploration, while ensuring that sequences generated at every optimization step remain near the learned viable mRNA manifold.

\begin{figure*}[!t]
\includegraphics[width=\linewidth]{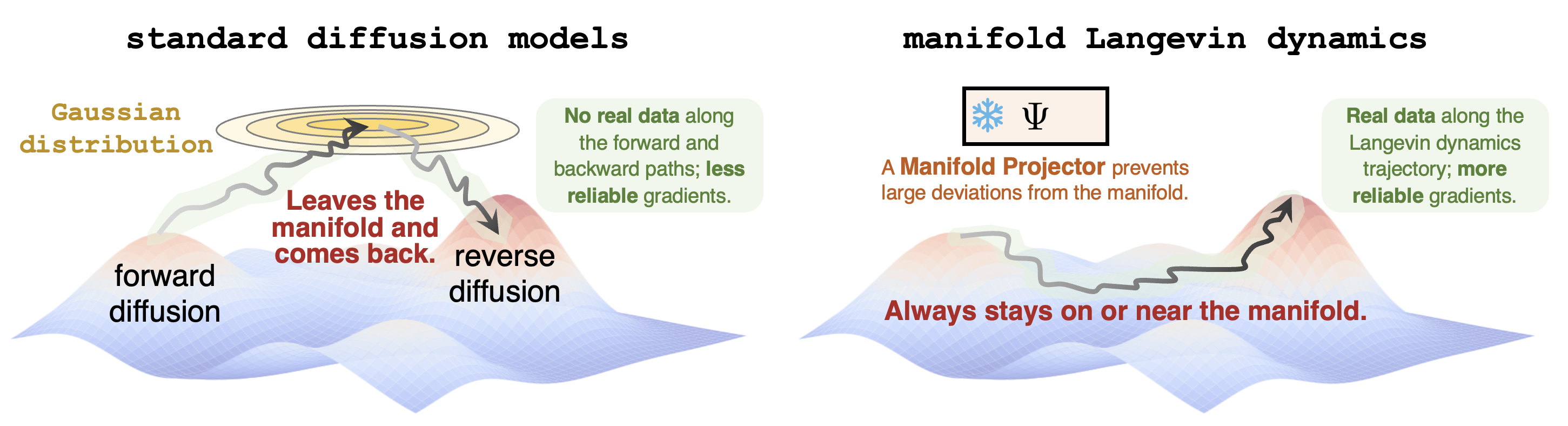}
\vspace{-16pt}
\caption{Illustration of manifold Langevin dynamics for optimization on the manifold of real data.}
\label{fig:teaser}
\vspace{-8pt}
\end{figure*}

Our choice to perform iterative optimization directly on the learned data manifold, rather than adopting diffusion-based generation from Gaussian noise, is motivated by several practical considerations. \textcolor{firebrick}{\textbf{First}}, mRNA sequence optimization is inherently local, where improvements are typically made relative to an existing sequence. \textcolor{firebrick}{\textbf{Second}}, property-guided Langevin dynamics rely on gradients from a predictor trained on real data, and therefore staying close to the data manifold helps obtain stable and meaningful guidance. In contrast, standard diffusion and flow-matching models traverse off-manifold regions during generation, which increases the risk of unreliable guidance from predictors. \textcolor{firebrick}{\textbf{Third}}, constraining the optimization trajectory to the manifold encourages the generated sequences to remain near regions supported by observed viable sequences, with the aim to promote biological plausibility. \textcolor{firebrick}{\textbf{Finally}}, refining an existing latent representation avoids a full noise-to-data generation process and supports efficient inference. Figure~\ref{fig:teaser} provides a conceptual illustration.

We evaluate \ours on three real-world mRNA datasets under diverse experimental settings. Across these datasets, \ours improves median property gain by up to 267\% over state-of-the-art methods and success rates by up to 31\%, while preserving biological viability proxies. In addition, \ours also achieves a competitive inference throughput among the evaluated methods.

\vspace{-4pt}
\section{Preliminaries}
\label{sec:preliminaries}
\vspace{-4pt}

\begin{figure*}[!ht]
\centering
\includegraphics[width=\textwidth]{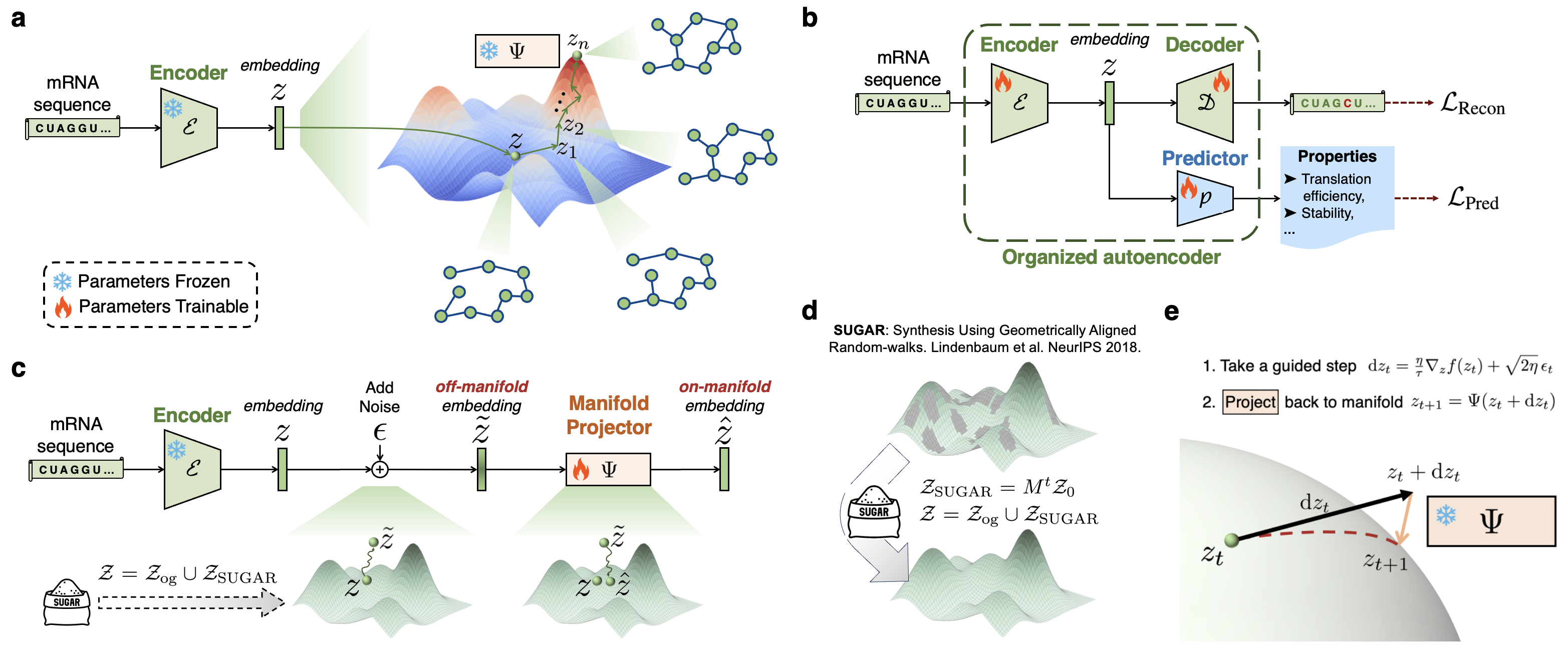}
\vspace{-12pt}
\caption{Schematic of \ours. \textbf{a,}~We encode a given input mRNA sequence into a latent manifold organized by property, and iteratively optimize the sequence along the property landscape. The manifold projector $\Psi$ encourages the optimization to stay on or near the manifold. Notably, we can decode the intermediate optimization result after every step. \textbf{b,}~The organized latent manifold is created by jointly optimizing reconstruction and property prediction objectives. \textbf{c,}~The manifold projector is trained to project perturbed data points back to the manifold. \textbf{d,}~For undersampled mRNA manifolds, we use SUGAR~\cite{SUGAR} to learn key dimensions in the manifold and fill undersampled regions. \textbf{e,}~During optimization, the manifold projector brings off-manifold points back to the manifold.}
\label{fig:schematic}
\vspace{-2pt}
\end{figure*}

\subsection{Manifold assumption and manifold learning}

\textbf{The manifold assumption}~\cite{ManifoldHypothesis1, ManifoldHypothesis2, ManifoldHypothesis3} is a fundamental principle in machine learning which hypothesizes that high-dimensional natural data lie near a low-dimensional manifold embedded in the ambient space~\cite{JiT}. Formally, each observation $x_i \in \mathbb{R}^n$ arises from a smooth nonlinear map $\mathbf{f} : \mathcal{M}^d \to \mathbb{R}^n$ applied to a latent variable $z_i \in \mathcal{M}^d$, where $d \ll n$.

\textbf{Manifold learning} methods seek to recover this latent structure by constructing representations that preserve intrinsic geometry~\cite{MAGIC, PHATE, MELD, CUTS, DSE, DiffKillR, ImageFlowNet, GAGA, Cflows, NeuralFIM}. A point is considered \emph{on-manifold} if it lies within the range of the nonlinear map $\mathbf{f}$, while \emph{off-manifold} points deviate from this structure and may correspond to invalid or adversarial samples~\cite{manifold_tangent_classifier, li2023discrete}. Thus, projecting updated points back to the manifold is critical for robustness and geometry-aware optimization~\cite{MPGD}.

\textbf{SUGAR}~\cite{SUGAR} is a diffusion geometry-based method that learns the intrinsic geometry of the data manifold and samples uniformly along the learned geometry. When observed data are limited, augmenting training sets with SUGAR enriches the learned manifold with geometry-preserving samples, particularly in sparse regions.

\textbf{Stochastic gradient descent~(SGD) on Riemannian manifolds}~\cite{StochasticGradientOnRiemannian} extends classical SGD by computing updates in the tangent space and mapping them back to the manifold via exponential maps or retractions. However, such methods assume an analytic form of the manifold. In contrast, the manifold underlying biological sequence space is not known in a closed form. \ours addresses this by directly \emph{learning} the projection operator, enabling optimization on data manifolds without requiring analytic solutions.

\vspace{-4pt}
\subsection{Langevin-dynamics and neural SDEs}
\vspace{-4pt}

\textbf{Langevin dynamics} is a stochastic process that describes motion driven by deterministic forces and thermal fluctuations. Its overdamped form with unit mobility is $\text{d}X_t = -\nabla U(X_t)\text{ d}t + \sqrt{2\tau} \text{ d}W_t$, where $U$ is the potential, $\tau$ is the temperature, and $W_t$ is Brownian motion. Under mild conditions, the process converges to the stationary distribution with $p \propto e^{-U/\tau}$, and its Euler discretization is often used for sampling from $p$~\cite{roberts1996exponential}. In generative modeling with score matching, a neural network $\boldsymbol{s_\theta}$ estimates log-density gradients of Gaussian-perturbed data for annealed Langevin sampling~\cite{Denoising_score_matching}.

\textbf{Neural stochastic differential equations~(neural SDEs)}~\cite{NeuralSDE} are SDEs whose drift and/or diffusion coefficients are parameterized by neural networks. They satisfy $\text{d}X_t = f_\theta(t, X_t)\text{ d}t + g_\theta(t, X_t)\text{ d}W_t$, where $f_\theta$ is the drift, $g_\theta$ is the diffusion coefficient, and $W_t$ is Brownian motion. A neural SDE has a direct connection with Langevin dynamics, as it recovers the Langevin equation above when its drift is $f_\theta = -\nabla U_\theta$ for a neural potential $U_\theta$ and its diffusion coefficient is $g_\theta = \sqrt{2\tau}I$.

\vspace{-4pt}
\section{\ours}
\vspace{-4pt}

The key components of our framework are \fancynumber{1}~\textbf{OAE}: an autoencoder jointly trained with a prediction model to learn a latent manifold of mRNA sequences organized by the target property~(Section~\ref{sec:OAE} and Figure~\ref{fig:schematic}b), \fancynumber{2}~\textbf{Manifold Projector}: a denoising autoencoder that brings the updated latent embeddings back to the learned data manifold during each step of optimization~(Section~\ref{sec:manifold_projector} and Figure~\ref{fig:schematic}c,~\ref{fig:schematic}e), and \fancynumber{3}~\textbf{Property-guided manifold Langevin dynamics}, a procedure that integrates the two aforementioned modules to enable property-guided generation~(Section~\ref{sec:PGOMLD} and Figure~\ref{fig:schematic}a). Once trained, these components allow \ours
to optimize the target property, such as translation efficiency or stability, of a given mRNA sequence.

\vspace{-4pt}
\subsection{Learning a latent manifold organized by property}
\label{sec:OAE}

We begin by training an \textbf{organized autoencoder~(OAE)}, where the latent manifold is implicitly structured via supervision from a property prediction task~(Figure~\ref{fig:schematic}b).

Similar to a vanilla autoencoder~\cite{AE}, the encoder $\mathcal{E}$ maps the input mRNA sequence $x$ to a latent representation $z$, which is decoded by $\mathcal{D}$ back to the sequence space. In addition to this standard architecture, a predictor $\mathcal{P}$ infers properties $\hat{y}$ from the embedding $z$. Formally, $z = \mathcal{E}(x), \,\, \hat{x} = \mathcal{D}(z), \,\, \hat{y} = \mathcal{P}(z)$. The latent manifold $\mathcal{Z}$ is thus shaped by jointly optimizing the reconstruction loss and the prediction loss~(Eqn~\eqref{eqn:oae_loss}), encouraging it to learn sequence-relevant information while being organized by the target properties. $\lambda_\text{Pred}$ and $\lambda_\text{Recon}$ are hyperparameters that balance between organizing the property landscape and capturing sequence information. They are empirically set as described in Supplementary Section~\ref{supp:hyperparameters}. Here, $x_i \in \mathbb{R}^{V}$ is the ground truth one-hot encoding of the nucleotide at position $i$ in sequence $x$ with an mRNA vocabulary of size $V$.
\begin{align}
\resizebox{0.92\textwidth}{!}{$
\begin{aligned}
    \mathcal{L}_\text{OAE} = \lambda_\text{Pred} \mathcal{L}_\text{Pred} + \lambda_\text{Recon} \mathcal{L}_\text{Recon} = \overbrace{\lambda_\text{Pred} \phantom{\frac{|}{|}} \mathbb{E}_{(x,y) \sim p_{\text{data}}} \|\hat{y} - y\|_2^2}^\text{optimize $\mathcal{P}$ and $\mathcal{E}$ with MSE} \hspace{6pt} - \overbrace{\lambda_\text{Recon} \phantom{\frac{|}{|}} \mathbb{E}_{x \sim p_{\text{data}}} \frac{1}{L}\sum_{i=1}^{L} \sum_{v=1}^{V} x_{i,v} \log \frac{\exp(\hat{x}_{i,v})}{\sum_{u=1}^{V}\exp(\hat{x}_{i,u})}}^\text{optimize $\mathcal{E}$ and $\mathcal{D}$ with CrossEntropy}
\label{eqn:oae_loss}
\end{aligned}
$}
\end{align}
\subsection{Learning a module to project updates toward the latent manifold}
\label{sec:manifold_projector}

\paragraph{Augmenting the data manifold}

mRNA datasets are typically scarce and unevenly sampled, which can hinder learning a faithful latent manifold directly from observed data. To mitigate this issue, we augment the latent embeddings using SUGAR~\cite{SUGAR}, a diffusion geometry-based method introduced in Section~\ref{sec:preliminaries}. This augmentation enriches the learned manifold with geometry-preserving samples, particularly in sparse regions, improving robustness of subsequent manifold projection.

Specifically, this preprocessing step yields an expanded latent set: $\mathcal{Z} = \mathcal{Z}_{\text{og}} \cup \mathcal{Z}_{\text{SUGAR}}, \hspace{4pt} \mathcal{Z}_{\text{SUGAR}} = M^t \mathcal{Z}_0,$
where $\mathcal{Z}_{\text{og}}$ are the original latent embeddings, $\mathcal{Z}_0$ are locally-sampled neighbors, and $M^t$ is the sparsity-corrected Markov diffusion transition matrix applied for $t$ steps. The SUGAR augmentation is performed in the latent space: we encode training sequences into latent representations, apply SUGAR to enrich this space with geometry-consistent neighbors, and then train the manifold projector.

\vspace{-4pt}
\paragraph{Training the manifold projector}

To access reliable property guidance during generation and keep the generated trajectories aligned with the latent data manifold, we introduce a manifold projector $\Psi$. As illustrated in Figure~\ref{fig:schematic}c and magnified in Figure~\ref{fig:schematic}e, $\Psi$ takes in a noisy optimized point $\tilde{z}$ and projects it back onto or near the manifold. To train $\Psi$, we construct a short Gaussian corruption chain from a clean embedding $z$ sampled from the SUGAR-augmented latent set: $\tilde z^{(0)}=z$ and $\tilde z^{(k)}=\tilde z^{(k-1)}+\sigma_k\epsilon_k$, where $\epsilon_k\overset{\mathrm{iid}}{\sim}\mathcal{N}(0,I)$ and $\sigma_k$ is the noise scale of the $k$-th corruption step. The projector learns to reverse each local corruption step, yielding the objective in Eqn~\eqref{eqn:manifold_projector}.
\vspace{-4pt}
\begin{equation}
\mathcal{L}_{\Psi} = \mathbb{E}_{z \sim p_{\mathcal{Z}},\,\epsilon_{1:K}} [ \sum_{k=1}^{K} \big\|\Psi(\tilde z^{(k)})-\tilde z^{(k-1)}\big\|_2^2 ]
\vspace{-4pt}
\label{eqn:manifold_projector}
\end{equation}
Here, $p_{\mathcal{Z}}$ is the empirical distribution of the augmented latent set. When $K=1$, this objective reduces to the standard denoising autoencoder loss~\cite{DAE}, as used in our single-step ablation. We keep $K$ small ($K=2$) to capture \textit{local denoising transitions near the data manifold}. Pseudocode and implementation details are provided in Supplementary Sections~\ref{supp:pseudo_code_mp} and~\ref{supp:hyperparameters}, respectively. At inference, the trained projector is applied once after each property-guided Langevin update.

\vspace{-4pt}
\subsection{Property-guided manifold Langevin dynamics}
\label{sec:PGOMLD}
\vspace{-4pt}

Next, we combine property-guided Langevin updates with the learned manifold projector to refine mRNA sequences. Given a trained encoder $\mathcal{E}$, a property predictor $\mathcal{P}$, and a manifold projector $\Psi$, our Langevin-dynamics framework optimizes sequences for a target property. Starting from the latent embedding $z = \mathcal{E}(x)$ of a sequence $x$, we iteratively update it using a gradient-based drift term $\nabla_z f(z)$, inject Gaussian noise $\epsilon$, and apply a manifold projector $\Psi(\cdot)$ to encourage proximity to the learned data manifold. We define the update rule as described in Eqn~\eqref{eqn:update_rule}.
\begin{equation}
z_{t+1} = \Psi\!\left(z_t + \text{d}z_t\right)
\qquad \text{d}z_t = \tfrac{\eta}{\tau}\nabla_z f(z_t) + \sqrt{2\eta}\,\epsilon_t, \quad \epsilon_t \sim \mathcal{N}(0, I)
\label{eqn:update_rule}
\end{equation}
Here, $\eta$ is the step size, $\tau$ is the temperature hyperparameter, and $\nabla_z f(z)$ denotes the property gradient given by the predictor $\mathcal{P}$. A smaller $\tau$ emphasizes focused updates along the gradient, while a larger $\tau$ encourages more diverse exploration. When $\tau \rightarrow \infty$, the property guidance vanishes and the update rule is dominated by the stochastic term, which becomes similar to generative modeling with the walkback algorithm~\cite{bengio2013generalizedDAE}. The manifold projector $\Psi$, analogous to the retraction in Riemannian SGD~\cite{StochasticGradientOnRiemannian}, is applied after each update to ensure that each step remains near the learned latent manifold, enabling interpretable and controllable generation trajectories.

\label{sec:method_optimize}
With the trained components $\mathcal{E}$, $\mathcal{P}$ and $\Psi$, we can optimize the target property of any given sequence. Notably, optimization entails both maximization and minimization: users can choose to increase or decrease the target property, depending on the application.

\section{Empirical Results}

We evaluate \ours on property-optimized mRNA sequence generation across multiple biologically relevant settings. The goal is to improve target properties such as translation efficiency and stability while preserving the biological viability of the generated sequences. To this end, we conducted comprehensive experiments to analyze the property gains, computational viability proxies, efficiency, and optimization trajectories.

\subsection{Experimental settings}

\paragraph{Datasets and the properties to optimize}
We evaluated \ours on three mRNA datasets that capture diverse contexts and experimental designs. The optimization objectives are italicized.
\begin{enumerate}[leftmargin=16pt, topsep=0pt, itemsep=4pt, parsep=0pt]
    \item \texttt{OpenVaccine} contains 2,400 mRNA sequences devised for COVID-19 mRNA vaccines, each with 107 nucleotides~\cite{Openvaccine}. The target property is mRNA \textit{stability}.
    \item \texttt{Zebrafish} consists of 5 subsets of zebrafish 5' UTR mRNAs each with 124 nucleotides, comprising more than 55,000 sequences in total~\cite{zebrafish_data}. The target property is \textit{translation efficiency}, which measures how effectively an mRNA sequence is translated into protein.
    \item \texttt{Ribosome} is a large-scale dataset of approximately 260,000 5' UTR mRNA sequences, each with 50 nucleotides~\cite{Optimus_5_prime}. The target property is the \textit{mean ribosome load}, defined as the average number of ribosomes bound to the mRNA during translation.
\end{enumerate}

\begin{figure*}[!tb]
\centering
\includegraphics[width=\textwidth]{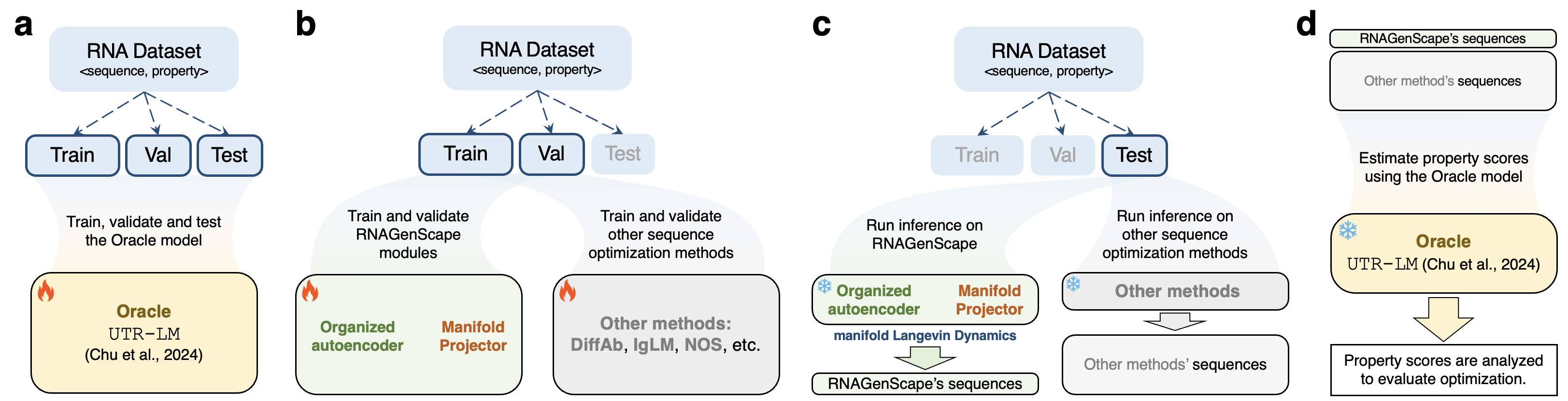}
\vspace{-12pt}
\caption{Overview of the experimental design. The same procedure is repeated for every mRNA dataset. \textbf{a,}~We first fine-tune and evaluate a pre-trained oracle model that predicts target properties from mRNA sequences. \textbf{b,}~We then train each sequence optimization method using the training and validation sets. \textbf{c,}~The trained methods are applied to optimize mRNA sequences starting from unseen test samples. \textbf{d,}~Finally, the oracle model is used to evaluate property improvement and success rate.}
\label{fig:experimental_setup}
\vspace{-6pt}
\end{figure*}

Our main experiments follow the design in Figure~\ref{fig:experimental_setup}. We train an oracle sequence-to-property model~(panel a), train each optimization method using only the train and validation data~(panel b), apply each method to improve sequences initialized from the unseen test set~(panel c), and report gains and success rates using the oracle so all methods being compared are scored fairly~(panel d). Throughout all evaluations, ``property gain/improvement'' refer to the oracle's predicted properties.

\paragraph{Baselines}
For direct comparison on the datasets above, we benchmark \ours against nine baselines for mRNA sequence optimization. DiffAb~\cite{DiffAb} performs sequence optimization using multinomial diffusion~\cite{hoogeboom2021argmax}. NOS-C~\cite{NOS} and NOS-D~\cite{NOS} perform property-guided sequence optimization using continuous and discrete transition distributions, respectively. IgLM~\cite{IgLM} refines candidates using an autoregressive language model. MFM~\cite{MFM} generalizes conditional flow matching by replacing Euclidean straight-line interpolants with approximate geodesic interpolants induced by a data-dependent Riemannian metric. gg-dWJS~\cite{gg_dWJS} follows gradient guidance from a discriminative property model, using discrete Markov chain to sample on a smoothed data manifold before jumping back to the discrete sequence space through one-step denoising. Persistent Contrastive Divergence~(PCD)~\cite{PCD} trains energy-based models by approximating the likelihood gradient using persistent Markov chains. Energy Matching~(EM)~\cite{EM} learns a time-independent scalar potential governing transport toward the data manifold and Langevin sampling. Manifold Preserving Guided Diffusion~(MPGD)~\cite{MPGD} guides reverse-diffusion sampling through updates constrained to manifold tangent spaces. All these models are trained from scratch on each of the three mRNA datasets with the correct interfaces with RNA $\texttt{A}$/$\texttt{U}$/$\texttt{G}$/$\texttt{C}$ nucleotides~(Figure~\ref{fig:experimental_setup}b), and then evaluated for mRNA sequence optimization~(Figure~\ref{fig:experimental_setup}c-d).

All experiments were repeated using five random seeds. Unless otherwise specified, results are reported as the mean $\pm$ standard deviation across the five runs. Experimental details and evaluation metrics are provided in Supplementary Sections~\ref{supp:hyperparameters} and~\ref{supp:eval_metrics}.

\begin{table*}[!tb]
\caption{Our proposed \ours generates mRNA sequences that achieve strong property improvement while maintaining favorable computational viability proxies. $\mathcal{M}$ denotes whether the method takes the manifold structure into consideration. In property optimization results, $\Delta$ denotes the median change in property and $\%$ denotes success rate~(percentage of mRNAs improved). In the biological viability checks, uORF OOF \texttt{AUG} denote out-of-frame start codon in the upstream open reading frame, Kozak similarity is the similarity to the consensus Kozak sequence, and mean pLDDT is a structural-confidence score averaged over nucleotides.}
\centering
\resizebox{\textwidth}{!}{%
    \begin{tabular}{lccccccccc}
    \toprule
    \multirow{3}{*}{Method} & \multirow{3}{*}{$\mathcal{M}$} & \multicolumn{4}{c}{\large \textbf{Property Optimization}} & \multicolumn{4}{c}{Biological Viability Checks} \\
    \cmidrule(lr){3-6} \cmidrule(lr){7-10}
    & & \multicolumn{2}{c}{$+$Property} & \multicolumn{2}{c}{$-$Property} & Initiation Interference & Initiation Context & Structural Stability & Structural Confidence \\
    \cmidrule(lr){3-4} \cmidrule(lr){5-6} \cmidrule(lr){7-7} \cmidrule(lr){8-8} \cmidrule(lr){9-9} \cmidrule(lr){10-10}
    & & $\Delta$~$\uparrow$ & \% $\uparrow$ & $\Delta$~$\downarrow$ & \% $\uparrow$ & uORF OOF \texttt{AUG} \%~$\downarrow$ & Kozak Similarity \%~$\uparrow$ & Minimum Free Energy~$\downarrow$ & Mean pLDDT~$\uparrow$ \\
    \midrule
    \multicolumn{8}{l}{\texttt{OpenVaccine~(n$\approx$2k)}, property to optimize: stability} \\
    \cmidrule(lr){1-10}
    \textcolor{gray}{Data in the test set} & \textcolor{gray}{---} & \textcolor{gray}{---} & \textcolor{gray}{---} & \textcolor{gray}{---} & \textcolor{gray}{---} & \textcolor{gray}{$1.75$ $\pm~0.12$} & \textcolor{gray}{$28.70$ $\pm~0.62$} & \textcolor{gray}{$-29.82$ $\pm~0.66$} & \textcolor{gray}{$42.18$ $\pm~0.24$} \\
    DiffAb~\cite{DiffAb} & \xmark & $+0.49$ \textcolor{gray}{$\pm~0.09$} & $72.5$ \textcolor{gray}{$\pm~3.4$} & $-0.20$ \textcolor{gray}{$\pm~0.16$} & $60.1$ \textcolor{gray}{$\pm~7.3$} & $1.52$ \textcolor{gray}{$\pm~0.55$} & $22.13$ \textcolor{gray}{$\pm~5.68$} & $-14.41$ \textcolor{gray}{$\pm~3.89$} & $35.57$ \textcolor{gray}{$\pm~3.00$} \\
    IgLM~\cite{IgLM} & \xmark & $+0.59$ \textcolor{gray}{$\pm~0.02$} & $80.6$ \textcolor{gray}{$\pm~0.9$} & $+0.43$ \textcolor{gray}{$\pm~0.06$} & $21.4$ \textcolor{gray}{$\pm~3.0$} & $2.01$ \textcolor{gray}{$\pm~0.16$} & $28.04$ \textcolor{gray}{$\pm~2.70$} & $-15.68$ \textcolor{gray}{$\pm~0.51$} & $38.00$ \textcolor{gray}{$\pm~0.77$} \\
    NOS-C~\cite{NOS} & \xmark & $+0.10$ \textcolor{gray}{$\pm~0.06$} & $61.4$ \textcolor{gray}{$\pm~6.7$} & $+0.03$ \textcolor{gray}{$\pm~0.00$} & $45.3$ \textcolor{gray}{$\pm~3.3$} & $1.77$ \textcolor{gray}{$\pm~0.22$} & $29.79$ \textcolor{gray}{$\pm~1.54$} & $-24.55$ \textcolor{gray}{$\pm~0.37$} & $39.63$ \textcolor{gray}{$\pm~0.35$} \\
    NOS-D~\cite{NOS} & \xmark & $+0.16$ \textcolor{gray}{$\pm~0.02$} & $72.2$ \textcolor{gray}{$\pm~1.5$} & $-0.06$ \textcolor{gray}{$\pm~0.02$} & $60.0$ \textcolor{gray}{$\pm~3.0$} & $1.73$ \textcolor{gray}{$\pm~0.23$} & $29.41$ \textcolor{gray}{$\pm~1.31$} & $-24.11$ \textcolor{gray}{$\pm~0.46$} & $39.94$ \textcolor{gray}{$\pm~0.70$} \\
    PCD~\cite{PCD} & \xmark & $+0.16$ \textcolor{gray}{$\pm~0.04$} & $56.1$ \textcolor{gray}{$\pm~0.4$} & $+0.16$ \textcolor{gray}{$\pm~0.05$} & $44.0$ \textcolor{gray}{$\pm~0.5$} & $2.67$ \textcolor{gray}{$\pm~0.13$} & $37.07$ \textcolor{gray}{$\pm~3.49$} & $-15.89$ \textcolor{gray}{$\pm~0.17$} & $32.12$ \textcolor{gray}{$\pm~0.84$} \\
    EM~\cite{EM} & \cmark & $+0.02$ \textcolor{gray}{$\pm~0.02$} & $54.2$ \textcolor{gray}{$\pm~5.0$} & $-0.03$ \textcolor{gray}{$\pm~0.02$} & $55.1$ \textcolor{gray}{$\pm~5.0$} & $1.69$ \textcolor{gray}{$\pm~0.25$} & $26.49$ \textcolor{gray}{$\pm~2.13$} & $-24.36$ \textcolor{gray}{$\pm~0.85$} & $38.15$ \textcolor{gray}{$\pm~0.47$} \\
    MFM~\cite{MFM} & \cmark & $+0.31$ \textcolor{gray}{$\pm~0.02$} & $69.7$ \textcolor{gray}{$\pm~1.7$} & $-0.21$ \textcolor{gray}{$\pm~0.07$} & $64.4$ \textcolor{gray}{$\pm~4.4$} & $2.08$ \textcolor{gray}{$\pm~0.29$} & $33.21$ \textcolor{gray}{$\pm~1.97$} & $-17.42$ \textcolor{gray}{$\pm~1.06$} & $34.10$ \textcolor{gray}{$\pm~0.29$} \\
    gg-dWJS~\cite{gg_dWJS} & \cmark & $+0.52$ \textcolor{gray}{$\pm~0.07$} & $70.4$ \textcolor{gray}{$\pm~5.2$} & $+0.34$ \textcolor{gray}{$\pm~0.02$} & $38.2$ \textcolor{gray}{$\pm~3.1$} & $\textbf{1.42}$ \textcolor{gray}{$\pm~0.45$} & $27.01$ \textcolor{gray}{$\pm~4.97$} & $-\phantom{0}9.11$ \textcolor{gray}{$\pm~4.94$} & $39.73$ \textcolor{gray}{$\pm~4.85$} \\
    MPGD~\cite{MPGD} & \cmark & $+0.20$ \textcolor{gray}{$\pm~0.05$} & $61.0$ \textcolor{gray}{$\pm~4.4$} & $-0.07$ \textcolor{gray}{$\pm~0.06$} & $53.9$ \textcolor{gray}{$\pm~3.9$} & $2.20$ \textcolor{gray}{$\pm~0.51$} & $31.22$ \textcolor{gray}{$\pm~4.01$} & $-19.08$ \textcolor{gray}{$\pm~1.08$} & $34.71$ \textcolor{gray}{$\pm~0.57$} \\
    \rowcolor{YaleBlue!12} \textbf{\ours~(ours)} & \cmark & $+\textbf{0.92}$ \textcolor{gray}{$\pm~0.22$} & $\textbf{83.6}$ \textcolor{gray}{$\pm~4.8$} & $-\textbf{0.77}$ \textcolor{gray}{$\pm~0.18$} & $\textbf{84.2}$ \textcolor{gray}{$\pm~2.2$} & $1.54$ \textcolor{gray}{$\pm~1.17$} & $\textbf{45.60}$ \textcolor{gray}{$\pm~9.21$} & $-\textbf{31.41}$ \textcolor{gray}{$\pm~5.17$} & $\textbf{43.97}$ \textcolor{gray}{$\pm~3.06$} \\
    Gain & --- & \textcolor{forestgreen}{$+56\%$} & \textcolor{forestgreen}{$+4\%$} & \textcolor{forestgreen}{$+267\%$} & \textcolor{forestgreen}{$+31\%$} & --- & --- & --- & --- \\

    \midrule
    \multicolumn{8}{l}{\texttt{Zebrafish~(n$\approx$55k)}, property to optimize: translation efficiency} \\
    \cmidrule(lr){1-10}

    \textcolor{gray}{Data in the test set} & \textcolor{gray}{---} & \textcolor{gray}{---} & \textcolor{gray}{---} & \textcolor{gray}{---} & \textcolor{gray}{---} & \textcolor{gray}{$2.38 \pm 0.06$} & \textcolor{gray}{$32.67 \pm 0.33$} & \textcolor{gray}{$-27.62 \pm 0.09$} & \textcolor{gray}{$30.42 \pm 0.30$} \\
    DiffAb~\cite{DiffAb} & \xmark & $+0.38$ \textcolor{gray}{$\pm~0.08$} & $61.4$ \textcolor{gray}{$\pm~3.0$} & $-0.79$ \textcolor{gray}{$\pm~0.08$} & $72.2$ \textcolor{gray}{$\pm~2.0$} & $3.57$ \textcolor{gray}{$\pm~0.17$} & $37.46$ \textcolor{gray}{$\pm~0.42$} & $-20.74$ \textcolor{gray}{$\pm~1.19$} & $31.73$ \textcolor{gray}{$\pm~0.45$}\\
    IgLM~\cite{IgLM} & \xmark & $-0.01$ \textcolor{gray}{$\pm~0.00$} & $49.8$ \textcolor{gray}{$\pm~0.1$} & $-0.26$ \textcolor{gray}{$\pm~0.04$} & $61.1$ \textcolor{gray}{$\pm~0.9$} & $2.72$ \textcolor{gray}{$\pm~0.09$} & $36.93$ \textcolor{gray}{$\pm~0.50$} & $-26.44$ \textcolor{gray}{$\pm~0.08$} & $30.19$ \textcolor{gray}{$\pm~0.41$} \\
    NOS-C~\cite{NOS} & \xmark & $+0.31$ \textcolor{gray}{$\pm~0.09$} & $63.5$ \textcolor{gray}{$\pm~7.2$} & $-0.14$ \textcolor{gray}{$\pm~0.03$} & $65.4$ \textcolor{gray}{$\pm~2.5$} & $2.24$ \textcolor{gray}{$\pm~0.09$} & $31.21$ \textcolor{gray}{$\pm~0.67$} & $-25.53$ \textcolor{gray}{$\pm~0.67$} & $31.26$ \textcolor{gray}{$\pm~1.18$} \\
    NOS-D~\cite{NOS} & \xmark & $+0.20$ \textcolor{gray}{$\pm~0.04$} & $65.9$ \textcolor{gray}{$\pm~4.5$} & $-0.23$ \textcolor{gray}{$\pm~0.04$} & $76.2$ \textcolor{gray}{$\pm~4.2$} & $2.28$ \textcolor{gray}{$\pm~0.04$} & $31.70$ \textcolor{gray}{$\pm~0.27$} & $-25.48$ \textcolor{gray}{$\pm~0.18$} & $30.98$ \textcolor{gray}{$\pm~0.38$} \\
    PCD~\cite{PCD} & \xmark & $-0.17$ \textcolor{gray}{$\pm~0.01$} & $46.0$ \textcolor{gray}{$\pm~0.7$} & $-0.18$ \textcolor{gray}{$\pm~0.01$} & $54.0$ \textcolor{gray}{$\pm~0.7$} & $2.68$ \textcolor{gray}{$\pm~0.04$} & $34.44$ \textcolor{gray}{$\pm~0.32$} & $-16.60$ \textcolor{gray}{$\pm~0.15$} & $32.07$ \textcolor{gray}{$\pm~0.51$} \\
    EM~\cite{EM} & \cmark & $+0.47$ \textcolor{gray}{$\pm~0.05$} & $69.1$ \textcolor{gray}{$\pm~3.2$} & $-0.73$ \textcolor{gray}{$\pm~0.13$} & $77.0$ \textcolor{gray}{$\pm~2.3$} & $1.89$ \textcolor{gray}{$\pm~0.04$} & $16.41$ \textcolor{gray}{$\pm~1.29$} & $-25.53$ \textcolor{gray}{$\pm~1.25$} & $30.94$ \textcolor{gray}{$\pm~0.15$} \\
    MFM~\cite{MFM} & \cmark & $+0.13$ \textcolor{gray}{$\pm~0.25$} & $55.0$ \textcolor{gray}{$\pm~8.1$} & $-0.36$ \textcolor{gray}{$\pm~0.42$} & $66.7$ \textcolor{gray}{$\pm~9.4$} & $2.30$ \textcolor{gray}{$\pm~0.16$} & $32.71$ \textcolor{gray}{$\pm~1.82$} & $-25.20$ \textcolor{gray}{$\pm~3.61$} & $31.64$ \textcolor{gray}{$\pm~0.10$} \\
    gg-dWJS~\cite{gg_dWJS} & \cmark & $-0.73$ \textcolor{gray}{$\pm~0.20$} & $32.1$ \textcolor{gray}{$\pm~5.1$} & $-1.14$ \textcolor{gray}{$\pm~0.26$} & $77.3$ \textcolor{gray}{$\pm~5.7$} & $3.60$ \textcolor{gray}{$\pm~0.37$} & $\textbf{40.68}$ \textcolor{gray}{$\pm~1.18$} & $-26.36$ \textcolor{gray}{$\pm~3.58$} & $30.81$ \textcolor{gray}{$\pm~0.88$} \\
    MPGD~\cite{MPGD} & \cmark & $-0.08$ \textcolor{gray}{$\pm~0.22$} & $47.5$ \textcolor{gray}{$\pm~7.2$} & $-0.90$ \textcolor{gray}{$\pm~0.20$} & $69.5$ \textcolor{gray}{$\pm~6.3$} & $2.15$ \textcolor{gray}{$\pm~0.41$} & $32.62$ \textcolor{gray}{$\pm~2.43$} & $-29.34$ \textcolor{gray}{$\pm~2.76$} & $32.54$ \textcolor{gray}{$\pm~1.92$} \\
    \rowcolor{YaleBlue!12} \textbf{\ours~(ours)} & \cmark & $+\textbf{0.79}$ \textcolor{gray}{$\pm~0.17$} & $\textbf{70.9}$ \textcolor{gray}{$\pm~5.0$} & $-\textbf{1.25}$ \textcolor{gray}{$\pm~0.53$} & $\textbf{77.3}$ \textcolor{gray}{$\pm~9.0$} & $\textbf{1.33}$ \textcolor{gray}{$\pm~0.15$} & $29.68$ \textcolor{gray}{$\pm~6.22$} & $-\textbf{29.55}$ \textcolor{gray}{$\pm~1.27$} & $\textbf{32.82}$ \textcolor{gray}{$\pm~1.82$} \\
    Gain & --- & \textcolor{forestgreen}{$+68\%$} & \textcolor{forestgreen}{$+3\%$} & \textcolor{forestgreen}{$+10\%$} & \textcolor{forestgreen}{$+0\%$} & --- & --- & --- & --- \\

    \midrule
    \multicolumn{8}{l}{\texttt{Ribosome~(n$\approx$260k)}, property to optimize: mean ribosome load} \\
    \cmidrule(lr){1-10}

    \textcolor{gray}{Data in the test set} & \textcolor{gray}{---} & \textcolor{gray}{---} & \textcolor{gray}{---} & \textcolor{gray}{---} & \textcolor{gray}{---} & \textcolor{gray}{$3.44 \pm 0.00$} & \textcolor{gray}{$19.31 \pm 0.00$} & \textcolor{gray}{$-\phantom{0}8.56 \pm 0.00$} & \textcolor{gray}{$41.99 \pm 0.00$} \\
    DiffAb~\cite{DiffAb} & \xmark & $+0.64$ \textcolor{gray}{$\pm~0.02$} & $86.2$ \textcolor{gray}{$\pm~1.5$} & $-1.10$ \textcolor{gray}{$\pm~0.05$} & $77.7$ \textcolor{gray}{$\pm~1.6$} & $\textbf{1.61}$ \textcolor{gray}{$\pm~0.21$} & $\phantom{0}8.24$ \textcolor{gray}{$\pm~0.99$} & $-\phantom{0}6.54$ \textcolor{gray}{$\pm~0.24$} & $40.85$ \textcolor{gray}{$\pm~0.19$} \\
    IgLM~\cite{IgLM} & \xmark & $+0.26$ \textcolor{gray}{$\pm~0.04$} & $66.1$ \textcolor{gray}{$\pm~1.1$} & $-0.42$ \textcolor{gray}{$\pm~0.02$} & $66.0$ \textcolor{gray}{$\pm~0.4$} & $2.37$ \textcolor{gray}{$\pm~0.04$} & $25.73$ \textcolor{gray}{$\pm~0.48$} & $-\phantom{0}8.20$ \textcolor{gray}{$\pm~0.09$} & $41.87$ \textcolor{gray}{$\pm~1.26$} \\
    NOS-C~\cite{NOS} & \xmark & $+0.08$ \textcolor{gray}{$\pm~0.01$} & $70.1$ \textcolor{gray}{$\pm~2.2$} & $-0.10$ \textcolor{gray}{$\pm~0.02$} & $60.5$ \textcolor{gray}{$\pm~3.0$} & $3.15$ \textcolor{gray}{$\pm~0.07$} & $18.47$ \textcolor{gray}{$\pm~0.37$} & $-\phantom{0}7.62$ \textcolor{gray}{$\pm~0.03$} & $41.03$ \textcolor{gray}{$\pm~0.96$} \\
    NOS-D~\cite{NOS} & \xmark & $+0.09$ \textcolor{gray}{$\pm~0.00$} & $74.8$ \textcolor{gray}{$\pm~1.0$} & $-0.12$ \textcolor{gray}{$\pm~0.01$} & $73.7$ \textcolor{gray}{$\pm~1.7$} & $3.14$ \textcolor{gray}{$\pm~0.08$} & $18.77$ \textcolor{gray}{$\pm~0.47$} & $-\phantom{0}7.13$ \textcolor{gray}{$\pm~0.05$} & $40.45$ \textcolor{gray}{$\pm~0.85$} \\
    PCD~\cite{PCD} & \xmark & $+0.03$ \textcolor{gray}{$\pm~0.02$} & $51.5$ \textcolor{gray}{$\pm~1.1$} & $+0.04$ \textcolor{gray}{$\pm~0.02$} & $48.5$ \textcolor{gray}{$\pm~0.9$} & $3.06$ \textcolor{gray}{$\pm~0.14$} & $18.43$ \textcolor{gray}{$\pm~1.46$} & $-\phantom{0}7.87$ \textcolor{gray}{$\pm~0.13$} & $42.70$ \textcolor{gray}{$\pm~0.94$} \\
    EM~\cite{EM} & \cmark & $+0.08$ \textcolor{gray}{$\pm~0.02$} & $68.6$ \textcolor{gray}{$\pm~4.9$} & $-0.01$ \textcolor{gray}{$\pm~0.02$} & $49.8$ \textcolor{gray}{$\pm~6.9$} & $2.64$ \textcolor{gray}{$\pm~0.15$} & $15.69$ \textcolor{gray}{$\pm~0.49$} & $-\phantom{0}8.24$ \textcolor{gray}{$\pm~0.21$} & $42.26$ \textcolor{gray}{$\pm~0.87$} \\
    MFM~\cite{MFM} & \cmark & $+0.01$ \textcolor{gray}{$\pm~0.02$} & $46.5$ \textcolor{gray}{$\pm~8.4$} & $-0.01$ \textcolor{gray}{$\pm~0.02$} & $49.2$ \textcolor{gray}{$\pm~9.2$} & $3.35$ \textcolor{gray}{$\pm~0.14$} & $19.98$ \textcolor{gray}{$\pm~0.27$} & $-\phantom{0}8.48$ \textcolor{gray}{$\pm~0.11$} & $41.81$ \textcolor{gray}{$\pm~0.57$} \\
    gg-dWJS~\cite{gg_dWJS} & \cmark & $+0.20$ \textcolor{gray}{$\pm~0.05$} & $59.0$ \textcolor{gray}{$\pm~1.6$} & $-0.07$ \textcolor{gray}{$\pm~0.05$} & $52.6$ \textcolor{gray}{$\pm~1.8$} & $2.87$ \textcolor{gray}{$\pm~0.10$} & $20.80$ \textcolor{gray}{$\pm~0.74$} & $-\phantom{0}7.38$ \textcolor{gray}{$\pm~0.44$} & $41.44$ \textcolor{gray}{$\pm~0.71$} \\
    MPGD~\cite{MPGD} & \cmark & $+0.60$ \textcolor{gray}{$\pm~0.03$} & $84.1$ \textcolor{gray}{$\pm~0.8$} & $-0.78$ \textcolor{gray}{$\pm~0.11$} & $90.7$ \textcolor{gray}{$\pm~5.1$} & $1.79$ \textcolor{gray}{$\pm~0.10$} & $19.75$ \textcolor{gray}{$\pm~0.83$} & $-\phantom{0}9.27$ \textcolor{gray}{$\pm~1.83$} & $41.45$ \textcolor{gray}{$\pm~0.31$} \\
    \rowcolor{YaleBlue!12} \textbf{\ours~(ours)} & \cmark & $+\textbf{0.93}$ \textcolor{gray}{$\pm~0.07$} & $\textbf{88.8}$ \textcolor{gray}{$\pm~3.0$} & $-\textbf{1.93}$ \textcolor{gray}{$\pm~0.57$} & $\textbf{92.8}$ \textcolor{gray}{$\pm~5.7$} & $3.35$ \textcolor{gray}{$\pm~0.78$} & $\textbf{36.46}$ \textcolor{gray}{$\pm~9.70$} & $-\textbf{10.17}$ \textcolor{gray}{$\pm~1.12$} & $\textbf{44.88}$ \textcolor{gray}{$\pm~5.59$} \\
    Gain & --- & \textcolor{forestgreen}{$+45\%$} & \textcolor{forestgreen}{$+3\%$} & \textcolor{forestgreen}{$+75\%$} & \textcolor{forestgreen}{$+2\%$} & --- & --- & --- & --- \\

    \bottomrule
    \end{tabular}
}
\vspace{-6pt}
\label{tab:main_result}
\end{table*}

\subsection{\ours generates property-optimized and computationally viable mRNA sequences}

Across all three datasets, we evaluate \ours along two complementary axes: improvement of the target property and preservation of biological viability proxies. Both aspects are jointly reported in Table~\ref{tab:main_result}. Property optimization is assessed using UTR-LM~\cite{UTR_LM}, a widely adopted mRNA language model, as the external oracle. UTR-LM is first pre-trained on over 480,000 5'~UTR sequences from diverse sources and subsequently fine-tuned for property prediction on each dataset~(Figure~\ref{fig:experimental_setup}a). After fine-tuning, the model is frozen and \textit{never accessed by any method} during their own training and inference~(Figure~\ref{fig:experimental_setup}b-c), ensuring an unbiased evaluation of the optimized sequences~(Figure~\ref{fig:experimental_setup}d).

In addition to analyzing the core property optimization performance, we apply a set of complementary checks grounded in established principles of mRNA translation to computationally assess the biological viability of optimized sequences. Specifically, we consider three classes of viability metrics. First, upstream out-of-frame start codons~(uORF OOF \texttt{AUG}) are known to suppress downstream protein expression and are therefore undesirable~\cite{uORF_OOF}. Second, Kozak similarity~\cite{kozak_similarity} measures how closely the initiation context matches the consensus Kozak sequence, with higher values indicating more favorable translation initiation. Third, we characterize RNA structure using minimum free energy~\cite{ViennaRNA} as a proxy for structural stability and mean predicted local distance difference test~(pLDDT)~\cite{AlphaFold2} computed with RhoFold~\cite{RhoFold} as a proxy for structural confidence. These checks provide supporting evidence of biological plausibility, but a more reliable assessment requires wet-lab validation.

\begin{wraptable}{r}{0.61\textwidth}
\vspace{-18pt}
\caption{Reproducing results in Table~\ref{tab:main_result} using third-party checkpoints which fine-tuned UTR-LM on external data as the oracles for translation efficiency (\href{https://huggingface.co/multimolecule/utrlm-te_el}{UTR-LM-TE}) and ribosome loading (\href{https://huggingface.co/multimolecule/utrlm-mrl}{UTR-LM-MRL}). These oracles are not fine-tuned on our datasets. Despite potential distribution shifts that limit direct comparison of predicted property values with those in Table~\ref{tab:main_result}, consistent gains suggest that our improvements are not solely due to shared predictor-oracle biases.}
\vspace{4pt}
\centering
\resizebox{\linewidth}{!}{%
    \begin{tabular}{lccccc}
    \toprule
    \multirow{3}{*}{Method} & \multirow{3}{*}{$\mathcal{M}$} & \multicolumn{4}{c}{\large \textbf{Property Optimization}}  \\
    \cmidrule(lr){3-6}
    & & \multicolumn{2}{c}{$+$Property} & \multicolumn{2}{c}{$-$Property} \\
    \cmidrule(lr){3-4} \cmidrule(lr){5-6}
    & & $\Delta$~$\uparrow$ & \% $\uparrow$ & $\Delta$~$\downarrow$ & \% $\uparrow$ \\

    \midrule
    \multicolumn{6}{l}{\texttt{Zebrafish~(n$\approx$55k)}, property to optimize: translation efficiency} \\
    \cmidrule(lr){1-6}
    DiffAb~\cite{DiffAb} & \xmark & $+0.25$ \textcolor{gray}{$\pm~0.04$} & $82.7$ \textcolor{gray}{$\pm~\phantom{0}3.0$} & $+0.00$ \textcolor{gray}{$\pm~0.03$} & $49.6$ \textcolor{gray}{$\pm~\phantom{0}4.4$} \\
    IgLM~\cite{IgLM} & \xmark & $+0.03$ \textcolor{gray}{$\pm~0.00$} & $56.5$ \textcolor{gray}{$\pm~\phantom{0}0.6$} & $-0.01$ \textcolor{gray}{$\pm~0.00$} & $52.5$ \textcolor{gray}{$\pm~\phantom{0}1.0$} \\
    NOS-C~\cite{NOS} & \xmark & $+0.07$ \textcolor{gray}{$\pm~0.02$} & $77.6$ \textcolor{gray}{$\pm~\phantom{0}8.4$} & $-0.04$ \textcolor{gray}{$\pm~0.04$} & $67.2$ \textcolor{gray}{$\pm~14.8$} \\
    NOS-D~\cite{NOS} & \xmark & $+0.06$ \textcolor{gray}{$\pm~0.00$} & $80.4$ \textcolor{gray}{$\pm~\phantom{0}1.2$} & $-0.12$ \textcolor{gray}{$\pm~0.02$} & $72.5$ \textcolor{gray}{$\pm~\phantom{0}4.4$} \\
    PCD~\cite{PCD} & \xmark & $-0.15$ \textcolor{gray}{$\pm~0.02$} & $29.4$ \textcolor{gray}{$\pm~\phantom{0}1.3$} & $-0.15$ \textcolor{gray}{$\pm~0.02$} & $70.6$ \textcolor{gray}{$\pm~\phantom{0}1.3$} \\
    EM~\cite{EM} & \cmark & $+0.06$ \textcolor{gray}{$\pm~0.05$} & $65.5$ \textcolor{gray}{$\pm~12.0$} & $-0.03$ \textcolor{gray}{$\pm~0.05$} & $56.4$ \textcolor{gray}{$\pm~12.9$} \\
    MFM~\cite{MFM} & \cmark & $+0.09$ \textcolor{gray}{$\pm~0.14$} & $62.6$ \textcolor{gray}{$\pm~24.4$} & $-0.04$ \textcolor{gray}{$\pm~0.11$} & $55.4$ \textcolor{gray}{$\pm~25.9$} \\
    gg-dWJS\cite{gg_dWJS} & \cmark & $+0.06$ \textcolor{gray}{$\pm~0.15$} & $55.7$ \textcolor{gray}{$\pm~15.8$} & $-0.10$ \textcolor{gray}{$\pm~0.17$} & $58.8$ \textcolor{gray}{$\pm~17.7$} \\
    MPGD~\cite{MPGD} & \cmark & $-0.08$ \textcolor{gray}{$\pm~0.10$} & $38.8$ \textcolor{gray}{$\pm~13.8$} & $-0.20$ \textcolor{gray}{$\pm~0.11$} & $75.8$ \textcolor{gray}{$\pm~12.5$} \\
    \rowcolor{YaleBlue!12} \textbf{\ours~(ours)} & \cmark & $+\textbf{0.40}$ \textcolor{gray}{$\pm~0.04$} & $\textbf{84.5}$ \textcolor{gray}{$\pm~\phantom{0}3.7$} & $-\textbf{0.25}$ \textcolor{gray}{$\pm~0.04$} & $\textbf{76.1}$ \textcolor{gray}{$\pm~\phantom{0}3.2$} \\

    \midrule
    \multicolumn{6}{l}{\texttt{Ribosome~(n$\approx$260k)}, property to optimize: mean ribosome load} \\
    \cmidrule(lr){1-6}
    DiffAb~\cite{DiffAb} & \xmark & $+0.50$ \textcolor{gray}{$\pm~0.05$} & $75.0$ \textcolor{gray}{$\pm~2.1$} & $-0.19$ \textcolor{gray}{$\pm~0.05$} & $59.5$ \textcolor{gray}{$\pm~1.7$} \\
    IgLM~\cite{IgLM} & \xmark & $+0.10$ \textcolor{gray}{$\pm~0.01$} & $55.5$ \textcolor{gray}{$\pm~0.3$} & $-0.08$ \textcolor{gray}{$\pm~0.01$} & $54.2$ \textcolor{gray}{$\pm~0.6$} \\
    NOS-C~\cite{NOS} & \xmark & $+0.20$ \textcolor{gray}{$\pm~0.02$} & $76.1$ \textcolor{gray}{$\pm~1.7$} & $+0.03$ \textcolor{gray}{$\pm~0.04$} & $45.9$ \textcolor{gray}{$\pm~4.6$} \\
    NOS-D~\cite{NOS} & \xmark & $+0.26$ \textcolor{gray}{$\pm~0.01$} & $88.5$ \textcolor{gray}{$\pm~0.9$} & $-0.27$ \textcolor{gray}{$\pm~0.01$} & $75.4$ \textcolor{gray}{$\pm~2.3$} \\
    PCD~\cite{PCD} & \xmark & $+0.01$ \textcolor{gray}{$\pm~0.04$} & $50.4$ \textcolor{gray}{$\pm~2.0$} & $-0.00$ \textcolor{gray}{$\pm~0.04$} & $49.9$ \textcolor{gray}{$\pm~1.9$} \\
    EM~\cite{EM} & \cmark & $+0.06$ \textcolor{gray}{$\pm~0.04$} & $61.1$ \textcolor{gray}{$\pm~7.6$} & $+0.01$ \textcolor{gray}{$\pm~0.04$} & $45.0$ \textcolor{gray}{$\pm~9.4$} \\
    MFM~\cite{MFM} & \cmark & $-0.00$ \textcolor{gray}{$\pm~0.01$} & $41.0$ \textcolor{gray}{$\pm~4.7$} & $-0.03$ \textcolor{gray}{$\pm~0.04$} & $52.1$ \textcolor{gray}{$\pm~9.2$} \\
    gg-dWJS\cite{gg_dWJS} & \cmark & $+0.29$ \textcolor{gray}{$\pm~0.13$} & $63.1$ \textcolor{gray}{$\pm~5.8$} & $+0.02$ \textcolor{gray}{$\pm~0.10$} & $49.0$ \textcolor{gray}{$\pm~4.4$} \\
    MPGD~\cite{MPGD} & \cmark & $+0.33$ \textcolor{gray}{$\pm~0.13$} & $63.9$ \textcolor{gray}{$\pm~6.2$} & $-0.63$ \textcolor{gray}{$\pm~0.27$} & $76.8$ \textcolor{gray}{$\pm~7.9$} \\
    \rowcolor{YaleBlue!12} \textbf{\ours~(ours)} & \cmark & $+\textbf{2.26}$ \textcolor{gray}{$\pm~0.47$} & $\textbf{96.8}$ \textcolor{gray}{$\pm~3.8$} & $-\textbf{0.65}$ \textcolor{gray}{$\pm~0.16$} & $\textbf{77.9}$ \textcolor{gray}{$\pm~4.3$} \\

    \bottomrule
    \end{tabular}
}
\vspace{-8pt}
\label{tab:supp_main_different_oracle}
\end{wraptable}

Across all three datasets and both optimization directions, \ours achieves the largest median property change and the highest success rate~(Table~\ref{tab:main_result}). Compared to the evaluated baselines, \ours attains up to a 267\% increase in median property improvement and up to a 31\% increase in success rate. In addition to better optimization performance, \ours generally produces sequences with relatively few aberrant upstream start codons, higher Kozak similarity, and more favorable minimum free energy profiles. Moreover, the viability metrics of \ours-generated sequences closely match those of real sequences in the unseen test set, indicating that substantial property gains can be achieved without compromising biological viability as evaluated using the computational viability proxies.

An additional evaluation with third-party oracles fine-tuned on disjoint external data also shows that \ours achieves the largest median property gains and success rate on Zebrafish and Ribosome in both optimization directions~(Table~\ref{tab:supp_main_different_oracle}), demonstrating robustness across evaluation oracles. See Supplementary Section~\ref{supp:eval_metrics} for details on these third-party oracles.

\subsection{Held-out validation on unseen near-optimal sequences}
\vspace{-4pt}

\begin{table}[!thb]
\caption{Mean normalized Hamming distance between generated sequences and their nearest favorable-property test-set references. Distances lie in $[0,1]$, with lower values being better.}
\vspace{-4pt}
\centering
\resizebox{\textwidth}{!}{%
    \begin{tabular}{cccccccccc>{\columncolor{YaleBlue!12}}c}
    \toprule
    & DiffAb & IgLM & NOS-C & NOS-D & PCD & EM & MFM & gg-dWJS & MPGD & \textbf{\ours} \\
    & \cite{DiffAb} & \cite{IgLM} & \cite{NOS} & \cite{NOS} & \cite{PCD} & \cite{EM} & \cite{MFM} & \cite{gg_dWJS} & \cite{MPGD} & \textbf{(ours)} \\
    \midrule
    \makecell[c]{\texttt{OpenVaccine}\\\texttt{(n$\approx$2k)}}
    & $0.424 \pm 0.015$ & $0.583 \pm 0.005$ & $0.421 \pm 0.001$ & $0.417 \pm 0.001$ & $0.546 \pm 0.001$ & $0.416 \pm 0.002$ & $0.433 \pm 0.009$ & $0.479 \pm 0.048$ & $0.412 \pm 0.002$ & $\textbf{0.411} \pm 0.023$ \\
    \cmidrule(l){1-11}
    \makecell[c]{\texttt{Zebrafish}\\\texttt{(n$\approx$55k)}}
    & $0.643 \pm 0.002$ & $0.642 \pm 0.001$ & $0.637 \pm 0.004$ & $0.635 \pm 0.004$ & $0.687 \pm 0.001$ & $0.653 \pm 0.004$ & $0.640 \pm 0.003$ & $0.638 \pm 0.003$ & $0.642 \pm 0.000$ & $\textbf{0.632} \pm 0.001$ \\
    \cmidrule(l){1-11}
    \makecell[c]{\texttt{Ribosome}\\\texttt{(n$\approx$260k)}}
    & $0.501 \pm 0.001$ & $0.509 \pm 0.000$ & $0.502 \pm 0.001$ & $0.497 \pm 0.001$ & $0.524 \pm 0.000$ & $0.510 \pm 0.002$ & $0.509 \pm 0.003$ & $0.504 \pm 0.002$ & $0.513 \pm 0.001$ & $\textbf{0.491} \pm 0.005$ \\
    \bottomrule
    \end{tabular}
}
\label{tab:held_out_validation}
\end{table}

As a supplementary consistency check, we assess whether generated candidates resemble held-out sequences with favorable oracle-predicted properties. We retain the standard train/validation/test splits and optimize sequences sampled from the test set. During evaluation, we select a subset from the test set whose properties are more than one standard deviation better than the population mean. This provides a curated set of favorable-property references unseen during training without changing the main experimental setup. Then, we report the mean normalized Hamming distance from generated candidates to their nearest sample in the favorable-property reference set.

\begin{table}[!tb]
\centering
\begin{minipage}{0.45\textwidth}
\centering
\caption{Reconstruction and prediction objectives in organized autoencoder are constructive and complementary.}
\resizebox{\textwidth}{!}{%
\begin{tabular}{ccccccc}
\toprule
& \multirow{2}{*}{$\mathcal{L}_\text{Recon}$} & \multirow{2}{*}{$\mathcal{L}_\text{Pred}$} & Reconstruction & \multicolumn{2}{c}{Property Prediction} \\
\cmidrule(lr){4-4} \cmidrule(lr){5-6}
&&& Nucleotide Acc.~$\uparrow$ & Pearson Corr.~$\uparrow$ & Spearman Corr.~$\uparrow$ \\
\midrule
\multirow{3}{*}{\makecell[c]{\texttt{OpenVaccine}\\\texttt{(n$\approx$2k)}}}
& \cmark & \xmark & \textbf{0.92} \textcolor{gray}{$\pm~0.00$} & 0.02 \textcolor{gray}{$\pm~0.08$} & 0.01 \textcolor{gray}{$\pm~0.08$} \\
& \xmark & \cmark & 0.14 \textcolor{gray}{$\pm~0.03$} & 0.65 \textcolor{gray}{$\pm~0.09$} & 0.72 \textcolor{gray}{$\pm~0.08$} \\
& \cellcolor{YaleBlue!12}\cmark & \cellcolor{YaleBlue!12}\cmark & \cellcolor{YaleBlue!12}\textbf{0.92} \textcolor{gray}{$\pm~0.01$} & \cellcolor{YaleBlue!12}\textbf{0.68} \textcolor{gray}{$\pm~0.06$} & \cellcolor{YaleBlue!12}\textbf{0.75} \textcolor{gray}{$\pm~0.04$} \\

\midrule
\multirow{3}{*}{\makecell[c]{\texttt{Zebrafish}\\\texttt{(n$\approx$55k)}}}
& \cmark & \xmark & \textbf{0.93} \textcolor{gray}{$\pm~0.01$} & 0.04 \textcolor{gray}{$\pm~0.01$} & 0.04 \textcolor{gray}{$\pm~0.02$} \\
& \xmark & \cmark & 0.13 \textcolor{gray}{$\pm~0.02$} & \textbf{0.76} \textcolor{gray}{$\pm~0.00$} & \textbf{0.75} \textcolor{gray}{$\pm~0.01$} \\
& \cellcolor{YaleBlue!12}\cmark & \cellcolor{YaleBlue!12}\cmark & \cellcolor{YaleBlue!12}0.89 \textcolor{gray}{$\pm~0.00$} & \cellcolor{YaleBlue!12}0.73 \textcolor{gray}{$\pm~0.01$} & \cellcolor{YaleBlue!12}0.73 \textcolor{gray}{$\pm~0.01$} \\

\midrule
\multirow{3}{*}{\makecell[c]{\texttt{Ribosome}\\\texttt{(n$\approx$260k)}}}
& \cmark & \xmark & \textbf{1.00} \textcolor{gray}{$\pm~0.00$} & 0.02 \textcolor{gray}{$\pm~0.01$} & 0.02 \textcolor{gray}{$\pm~0.01$} \\
& \xmark & \cmark & 0.15 \textcolor{gray}{$\pm~0.02$} & \textbf{0.92} \textcolor{gray}{$\pm~0.00$} & \textbf{0.90} \textcolor{gray}{$\pm~0.00$} \\
& \cellcolor{YaleBlue!12}\cmark & \cellcolor{YaleBlue!12}\cmark & \cellcolor{YaleBlue!12}\textbf{1.00} \textcolor{gray}{$\pm~0.00$} & \cellcolor{YaleBlue!12}\textbf{0.92} \textcolor{gray}{$\pm~0.00$} & \cellcolor{YaleBlue!12}\textbf{0.90} \textcolor{gray}{$\pm~0.00$} \\
\bottomrule
\end{tabular}
}
\label{tab:OAE_predictor}
\end{minipage}
\hfill
\begin{minipage}{0.5\textwidth}
\centering
\caption{The manifold projector $\Psi$ substantially improves property optimization by enforcing optimization along the data manifold.}
\resizebox{\textwidth}{!}{%
\begin{tabular}{lccccc}
\toprule
& \multirow{2}{*}{$\Psi$} & \multicolumn{2}{c}{$+$property} & \multicolumn{2}{c}{$-$property} \\
\cmidrule(lr){3-4} \cmidrule(lr){5-6}
& & $\Delta$ $\uparrow$ & \% $\uparrow$ & $\Delta$ $\downarrow$ & \% $\uparrow$ \\
\midrule
\multirow{2}{*}{\makecell[c]{\texttt{OpenVaccine}\\\texttt{(n$\approx$2k)}}}
& \xmark & $0.26$ \textcolor{gray}{$\pm\,0.09$} & $63.5$ \textcolor{gray}{$\pm\,1.2$} & $-0.69$ \textcolor{gray}{$\pm\,0.17$} & $79.6$ \textcolor{gray}{$\pm\,2.0$} \\

& \cellcolor{YaleBlue!12}\cmark & \cellcolor{YaleBlue!12}$\textbf{0.92}$ \textcolor{gray}{$\pm\,0.22$} & \cellcolor{YaleBlue!12}$\textbf{83.6}$ \textcolor{gray}{$\pm\,4.8$} & \cellcolor{YaleBlue!12}$-\textbf{0.77}$ \textcolor{gray}{$\pm\,0.18$} & \cellcolor{YaleBlue!12}$\textbf{84.2}$ \textcolor{gray}{$\pm\,2.2$} \\

\midrule
\multirow{2}{*}{\makecell[c]{\texttt{Zebrafish}\\\texttt{(n$\approx$55k)}}}
& \xmark & $0.77$ \textcolor{gray}{$\pm\,0.25$} & $68.3$ \textcolor{gray}{$\pm\,8.9$} & $-1.05$ \textcolor{gray}{$\pm\,0.48$} & $75.0$ \textcolor{gray}{$\pm\,9.7$} \\

& \cellcolor{YaleBlue!12}\cmark & \cellcolor{YaleBlue!12}$\textbf{0.79}$ \textcolor{gray}{$\pm\,0.17$} & \cellcolor{YaleBlue!12}$\textbf{70.9}$ \textcolor{gray}{$\pm\,5.0$} & \cellcolor{YaleBlue!12}$-\textbf{1.25}$ \textcolor{gray}{$\pm\,0.53$} & \cellcolor{YaleBlue!12}$\textbf{77.3}$ \textcolor{gray}{$\pm\,9.0$} \\

\midrule
\multirow{2}{*}{\makecell[c]{\texttt{Ribosome}\\\texttt{(n$\approx$260k)}}}
& \xmark & $0.58$ \textcolor{gray}{$\pm\,0.12$} & $81.1$ \textcolor{gray}{$\pm\,6.9$} & $-0.58$ \textcolor{gray}{$\pm\,0.24$} & $67.1$ \textcolor{gray}{$\pm\,6.9$} \\
& \cellcolor{YaleBlue!12}\cmark & \cellcolor{YaleBlue!12}$\textbf{0.93}$ \textcolor{gray}{$\pm\,0.07$} & \cellcolor{YaleBlue!12}$\textbf{88.8}$ \textcolor{gray}{$\pm\,3.0$} & \cellcolor{YaleBlue!12}$-\textbf{1.93}$ \textcolor{gray}{$\pm\,0.57$} & \cellcolor{YaleBlue!12}$\textbf{92.8}$ \textcolor{gray}{$\pm\,5.7$} \\
\bottomrule
\end{tabular}
}
\label{tab:ablation_manifold_projector}
\end{minipage}
\end{table}

\begin{table}[!tb]
\caption{Ablation study on the data augmentation ratio of SUGAR.}
\label{supp:ablation_sugar}
\centering
\resizebox{0.9\textwidth}{!}{%
    \begin{tabular}{lcccccccccccc}
    \toprule
    & \multicolumn{4}{c}{\texttt{OpenVaccine (n$\approx$2k)}} & \multicolumn{4}{c}{\texttt{Zebrafish (n$\approx$55k)}} & \multicolumn{4}{c}{\texttt{Ribosome-loading (n$\approx$260k)}} \\
    \cmidrule(lr){2-5} \cmidrule(lr){6-9} \cmidrule(lr){10-13}
    & \multicolumn{2}{c}{$+$property} & \multicolumn{2}{c}{$-$property}
    & \multicolumn{2}{c}{$+$property} & \multicolumn{2}{c}{$-$property}
    & \multicolumn{2}{c}{$+$property} & \multicolumn{2}{c}{$-$property} \\
    \cmidrule(lr){2-3} \cmidrule(lr){4-5} \cmidrule(lr){6-7} \cmidrule(lr){8-9} \cmidrule(lr){10-11} \cmidrule(lr){12-13}
    SUGAR ratio
    & $\Delta$ $\uparrow$ & \% $\uparrow$ & $\Delta$ $\downarrow$ & \% $\uparrow$
    & $\Delta$ $\uparrow$ & \% $\uparrow$ & $\Delta$ $\downarrow$ & \% $\uparrow$
    & $\Delta$ $\uparrow$ & \% $\uparrow$ & $\Delta$ $\downarrow$ & \% $\uparrow$ \\
    \midrule
    0 & $+0.44$ & $73.2$ & $-0.97$ & $\textbf{86.2}$ & $+0.14$ & $54.7$ & $-0.98$ & $73.0$ & $+0.83$ & $83.7$ & $-0.74$ & $67.9$ \\
    0.01 & $+0.08$ & $56.9$ & $-0.62$ & $67.5$ & $+0.69$ & $70.7$ & $-1.20$ & $\textbf{80.0}$ & $+0.53$ & $78.3$ & $-0.20$ & $51.3$ \\
    0.1 & $+0.20$ & $58.1$ & $-0.51$ & $74.0$ & $+0.65$ & $70.8$ & $-0.93$ & $75.3$ & $+0.49$ & $69.6$ & $-1.18$ & $71.4$ \\
    \rowcolor{YaleBlue!12} 1.0 & $+\textbf{0.92}$ & $\textbf{83.6}$ & $-0.77$ & $84.2$ & $+\textbf{0.79}$ & $\textbf{70.9}$ & $-\textbf{1.25}$ & $77.3$ & $+\textbf{0.93}$ & $\textbf{88.8}$ & $-\textbf{1.93}$ & $\textbf{92.8}$ \\
    10.0 & $-0.13$ & $57.1$ & $-\textbf{1.18}$ & $80.8$ & $-0.07$ & $47.9$ & $-1.10$ & $78.9$ & $+0.29$ & $63.1$ & $-1.44$ & $82.3$ \\
    \bottomrule
    \end{tabular}
}
\end{table}

\begin{wrapfigure}[20]{r}{0.5\textwidth}
\vspace{-6pt}
\includegraphics[width=\linewidth]{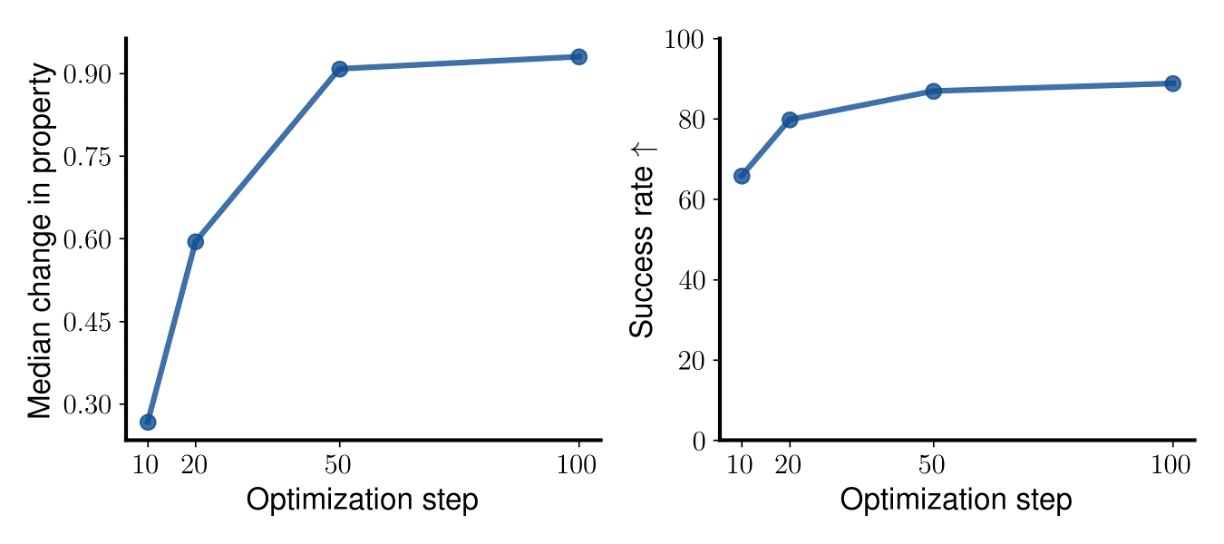}
\vspace{-20pt}
\caption{\ours optimization is step-efficient and remains stable over a range of optimization steps.}
\label{fig:ablation_step}
\vspace{4pt}
\includegraphics[width=\linewidth]{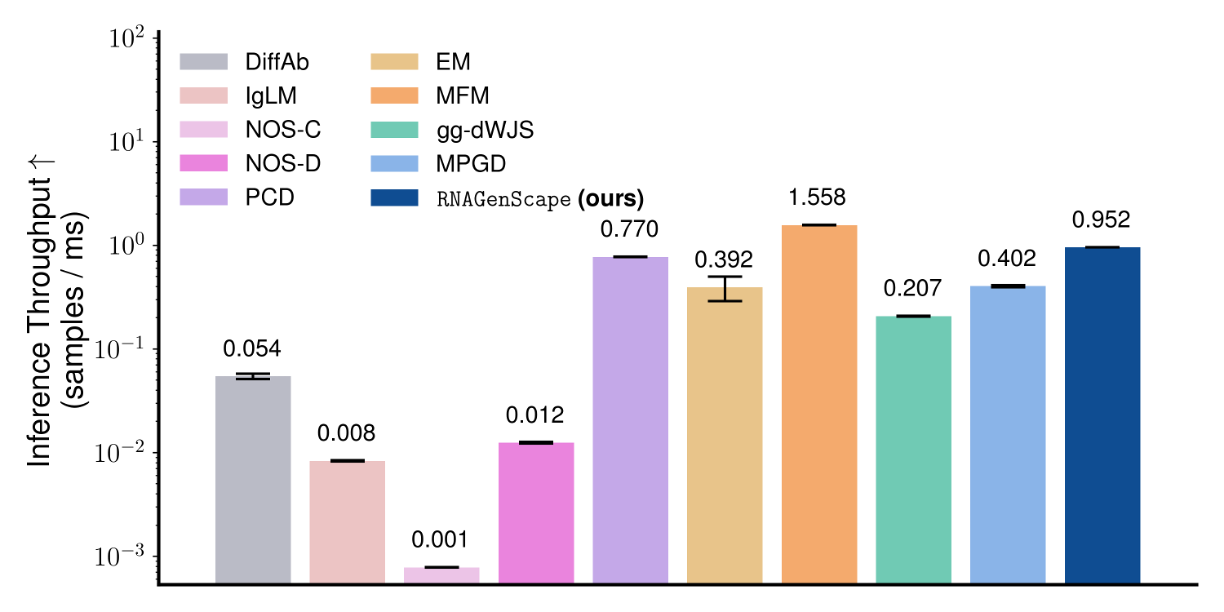}
\vspace{-20pt}
\caption{Inference throughput measured using CUDA-synchronized wall-clock time.}
\label{fig:efficiency}
\end{wrapfigure}

Table~\ref{tab:held_out_validation} reports these distances. Across all datasets, \ours consistently achieves the lowest Hamming distance compared to alternative biological sequence generation methods. This consistency check provides complementary and indirect evidence that generated candidates resemble real sequences with favorable properties.

\subsection{Ablation studies}
\label{sec:ablation}

\paragraph{Organized autoencoder}

Since the optimization process in \ours is performed in the learned latent manifold, we verify that the OAE provides both accurate reconstruction and property prediction~(Table~\ref{tab:OAE_predictor}). Additionally, we demonstrate that the reconstruction and prediction objectives~($\mathcal{L}_\text{Recon}$ and $\mathcal{L}_\text{Pred}$) are generally constructive to each other, which justifies our OAE design.

\paragraph{Manifold projector}
Next, we show that the manifold projector $\Psi$ is essential for the optimization performance of \ours~(Table~\ref{tab:ablation_manifold_projector}). Without $\Psi$, optimization becomes substantially less effective across datasets. This behavior is expected: updating latent representations solely by following the property gradient allows trajectories to drift away from the learned data manifold, entering regions that are weakly supported by real sequences. In these off-manifold regions, the predictor is less constrained by training data, which may contribute to diminished optimization gains. Further analysis of predictor behavior in different latent regions and ablations of projector design as well as reconstruction weight are provided in Supplementary Sections~\ref{supp:property_gradient} and~\ref{supp:additional_ablation}, respectively.

\paragraph{SUGAR augmentation}
Table~\ref{supp:ablation_sugar} shows that the default SUGAR ratio improves property optimization in most settings, while excessive augmentation can reduce performance.

\paragraph{Optimization steps}
We further analyze how sensitive \ours is to the number of optimization steps. The results in Figure~\ref{fig:ablation_step} show that our method is able to converge quickly and remains stable over a range of optimization steps during Langevin dynamics updates.

\paragraph{\textit{De novo} generative methods and classic optimization algorithms}
The results in Supplementary Section~\ref{supp:comparison_de_novo} show that~(1) \textit{de novo} generative models exhibit limited property control and~(2) classic optimization strategies in the OAE latent manifold also underperform \ours, indicating that our manifold Langevin dynamics contributed substantially to the favorable performance.

\vspace{-4pt}
\subsection{Efficiency and scalability}
\label{sec:results_efficiency}
\vspace{-4pt}

In addition to its other merits, \ours is also highly efficient at inference time. As reported in Figure~\ref{fig:efficiency}, it achieves a throughput of 0.952 samples/ms, second only to MFM at 1.558 samples/ms. This efficiency makes \ours well suited for large-scale or iterative design workflows where high throughput is essential.

\vspace{-4pt}
\subsection{Case study: \ours produces structured, data-aligned optimization trajectories}
\label{sec:results_trajectories}
\vspace{-4pt}

\begin{figure*}[!tb]
\centering
\includegraphics[width=0.95\textwidth]{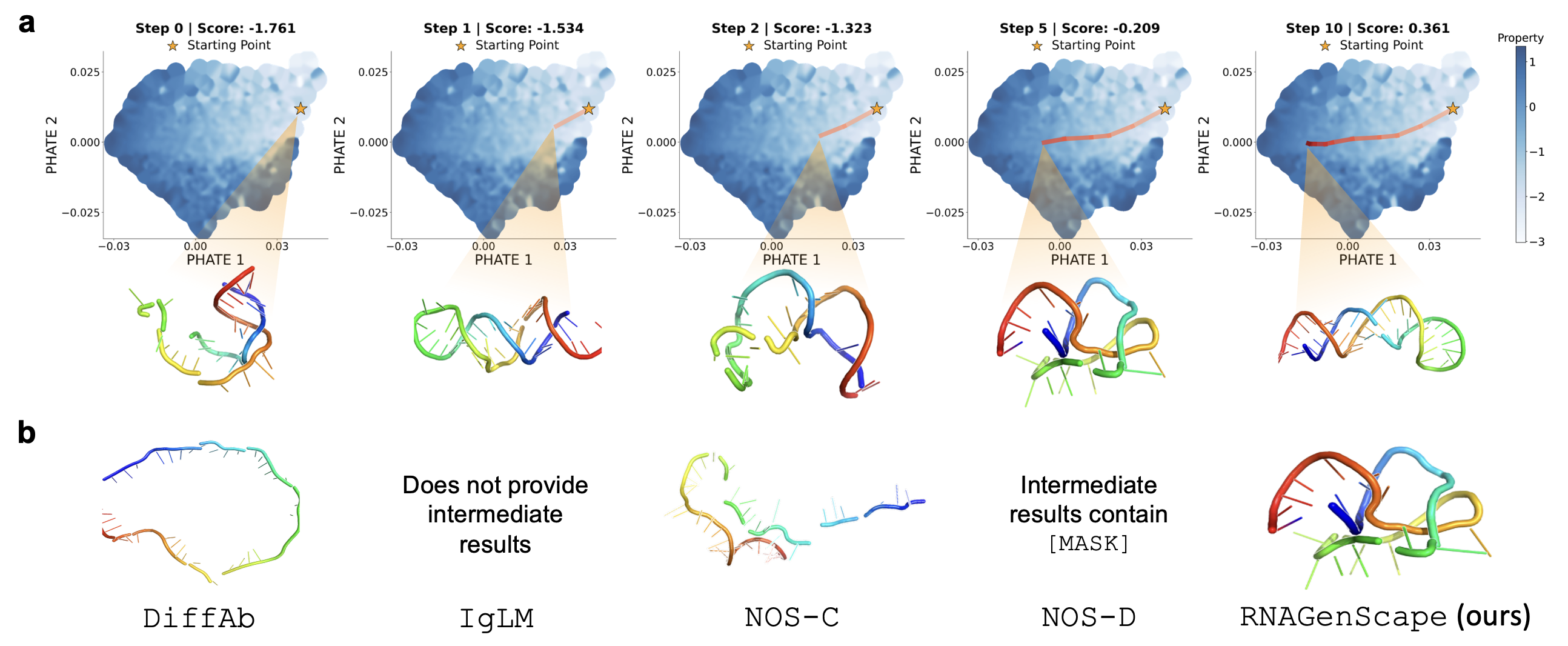}
\caption{A sample latent space trajectory of \ours over 10 optimization steps. The trajectory follows a smooth path with steady improvement in the target property. \textbf{a,}~The trajectory in the PHATE space and the corresponding 3D structures. \textbf{b,}~Intermediate results of \ours and competing methods midway through property optimization~(at the 5\textsuperscript{th} step out of 10 steps).}
\label{fig:latent_space_trajs}
\vspace{-12pt}
\end{figure*}

To illustrate how \ours iteratively navigates along a learned latent manifold, we visualize a sample optimization run in Figure~\ref{fig:latent_space_trajs}a. The trajectory exhibits monotonic increases in the target property, while remaining near regions populated by real sequences. Supplementary Sections~\ref{supp:additional_trajectories} and~\ref{sec:method_interpolate} provide further trajectories and sequence interpolation results, respectively. These trajectories are direct consequences of the manifold-constrained dynamics, which guided each step toward high-property regions while staying on the manifold. As a qualitative illustration of biological plausibility, we show that the intermediate results of \ours optimization can be properly folded by RhoFold~\cite{RhoFold} into 3D structures~(Figure~\ref{fig:latent_space_trajs}b). In contrast, folding intermediate products of several other methods causes failure, as indicated by fragmented structures~(Figure~\ref{fig:latent_space_trajs}b).

\begin{figure*}[!th]
\centering
\includegraphics[width=\textwidth]{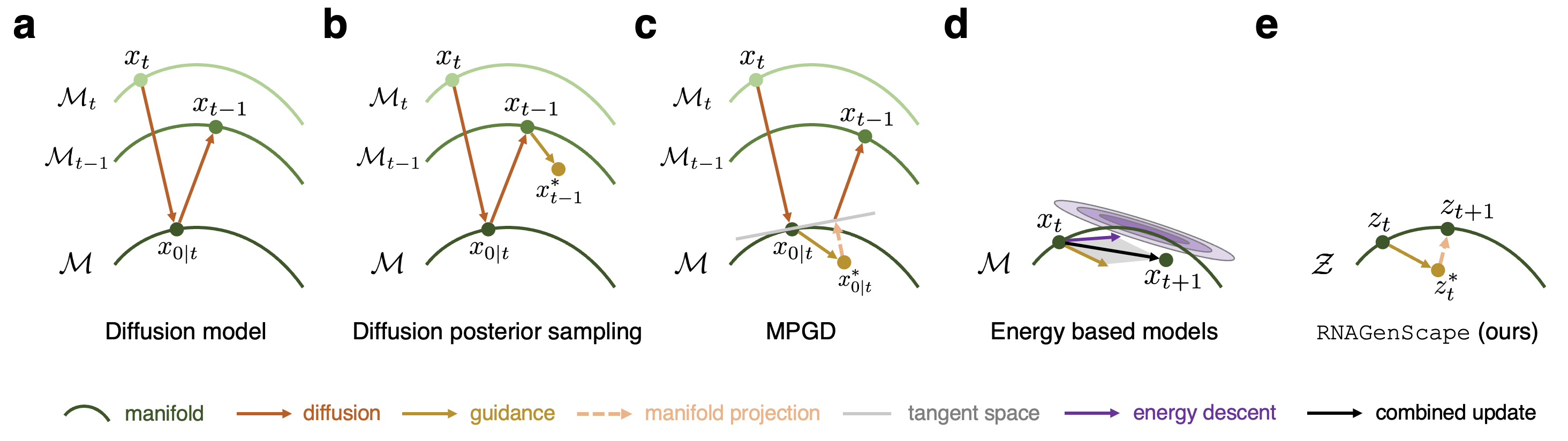}
\vspace{-12pt}
\caption{Conceptual comparison of sampling and guidance mechanisms relevant to our property-guided optimization problem. \textbf{a,} Unguided reverse diffusion. \textbf{b,} Diffusion posterior sampling~\cite{DSP} adds guidance to the reverse diffusion. \textbf{c,} Manifold Preserving Guided Diffusion~(MPGD)~\cite{MPGD} constrains guidance through a local manifold projection before reverse diffusion. \textbf{d,} Guided Langevin sampling in energy-based models~\cite{EM} combine guidance with energy descent; contours depict the learned energy landscape. \textbf{e,} \ours alternates property-guided updates and a SUGAR-assisted denoising projection in the OAE latent space. The distinct projection mechanisms used by MPGD and \ours are represented in the same color since they share the same high-level concept, whereas implementation details and stochastic perturbations are omitted from the illustration.}
\label{fig:conceptual_method_comparison}
\end{figure*}

\vspace{-4pt}
\section{Related Works}
\vspace{-4pt}

\paragraph{Sequence-to-function modeling}
Predicting quantitative properties such as expression and stability from biological sequences is central to sequence design~\cite{DNA_to_function}. Models trained on high-throughput experimental data capture sequence-function relationships in regulatory regions such as 5'~UTRs and promoters~\cite{Optimus_5_prime, vaishnav2022evolution}. ConvNets~\cite{chen2024mrna} and Transformers~\cite{RNAdegformer} capture dependencies in mRNA sequences and provide property predictors for downstream optimization.

\paragraph{Generative models for design}
\textit{De novo} generative models create novel biological sequences using VAEs~\cite{VAE, sinai2017variational, castillo2024optimizing}, adversarial models~\cite{GAN, prykhodko2019novo, mendez2020novo, ProteinGAN}, autoregressive language models~\cite{ProGen}, and diffusion models~\cite{DDPM, RFdiffusion}. ProteinMPNN~\cite{ProteinMPNN} generates sequences conditioned on protein structures, while RFdiffusion~\cite{RFdiffusion} generates protein structures with or without design constraints. These approaches provide foundations for biological design; our setting emphasizes property-guided optimization of an existing mRNA sequence.

\paragraph{Optimization of biological sequences}
Classic gradient-based and sampling-based optimization strategies~\cite{GradientAscent, zinkevich2003online, brooks1998markov, MCMC_ML, selman2006hill} can improve known sequences, but do not inherently constrain intermediate variants to the biological data distribution. Differentiable sequence optimization uses continuous relaxations~\cite{linder2019deep, FastSeqProp} or learned latent representations, as in ReLSO~\cite{ReLSO}. Other approaches use reinforcement learning~\cite{eastman2018solving}, Monte Carlo search~\cite{DesiRNA}, or multi-objective Bayesian optimization~\cite{LaMBO}. Deep generative methods such as DiffAb~\cite{DiffAb}, IgLM~\cite{IgLM}, and NOS~\cite{NOS} further connect sequence generation with design objectives. Our framework combines property-guided optimization with a learned manifold correction at each optimization step.

\paragraph{Energy-based and manifold-guided methods}
Energy-based and manifold-aware methods provide precedents for balancing property improvement with consistency with a learned data distribution. Models trained with Persistent Contrastive Divergence~(PCD)~\cite{PCD} and Energy Matching~(EM)~\cite{EM} provide energy priors that can be combined with property objectives. Metric Flow Matching~(MFM)~\cite{MFM} incorporates geometry into transport paths, while gradient-guided discrete walk-jump sampling~(gg-dWJS)~\cite{gg_dWJS} couples property guidance with sampling and denoising in a smoothed sequence space. Manifold Preserving Guided Diffusion~(MPGD)~\cite{MPGD} addresses guidance-induced departures from the manifold within reverse diffusion. \ours shares these motivations but combines property-organized mRNA embeddings with a denoising map trained on SUGAR-augmented embeddings, applied after each property-guided Langevin update without diffusion. Our contribution is an mRNA optimization framework with manifold learning and projection mechanisms distinct from those in prior works, evaluated computationally on real-world datasets of varying sizes. Figure~\ref{fig:conceptual_method_comparison} illustrates the mechanistic distinctions.

\paragraph{Integration of structural context}
Structural context provides additional information for biological sequence modeling. ProteinMPNN~\cite{ProteinMPNN} and diffusion-based inverse folding~\cite{yi2023graph} condition sequence generation on 3D structures. ImmunoStruct~\cite{ImmunoStruct} jointly models protein sequence, structure, and biochemical properties to predict immunogenicity. CellSpliceNet~\cite{CellSpliceNet} integrates sequence, secondary structure, and gene expression to predict alternative splicing. EternaFold~\cite{EternaFold} also models RNA secondary structure. Incorporating structural context is a potential extension to \ours.

\vspace{-4pt}
\section{Conclusion}
\vspace{-2pt}

We introduced \ours, a manifold Langevin dynamics framework for property-optimized mRNA sequence generation. Rather than optimizing in the ambient sequence space, \ours operates directly on a learned manifold of viable mRNA sequences. In this setting, property gradients are more reliable, biological viability is better preserved, and optimization can be performed with lower computational cost. Comprehensive experiments on three real-world mRNA datasets show that our method leads to greater property gains and higher success rates under realistic data regimes. Limitations, future work, and potential impacts are discussed in Supplementary Sections~\ref{supp:limitations} and~\ref{supp:impact}.

\section*{Acknowledgements}
S.K. is funded by the National Institutes of Health~(NIGMSR01GM135929, R01GM130847), the National Science Foundation (NSF CAREER award IIS-2047856, NSF IIS-2403317, NSF DMS-2327211, NSF CISE-2403317) and the National Cancer Institute (1OT2CA319802-01). S.K is also funded by the Sloan Fellowship FG-2021-15883 and the Novo Nordisk grant GR112933. A.J.G. is funded by the National Institutes of Health~(R01HD100035, R35GM122580).


\bibliography{references}
\bibliographystyle{unsrt}

\renewcommand{\thefigure}{S\arabic{figure}}
\renewcommand{\theHfigure}{S\arabic{figure}}
\setcounter{figure}{0}
\renewcommand{\thetable}{S\arabic{table}}
\renewcommand{\theHtable}{S\arabic{table}}
\setcounter{table}{0}

\clearpage
\newpage

\renewcommand\appendixpagename{\centering\noindent\rule{\textwidth}{2pt} \LARGE Technical Appendices \\ \normalsize \noindent\rule{\textwidth}{1pt}}

\begin{appendices}

\appendix
\onecolumn
\appendixpage

\begingroup
\setstretch{0}
\setcounter{tocdepth}{2}
\tableofcontents
\endgroup

\addtocontents{toc}{\protect\setcounter{tocdepth}{2}}

\vspace{8pt}

\section{Pseudocode of the manifold projector}
\label{supp:pseudo_code_mp}
\definecolor{codetext}{rgb}{0.25,0.5,0.5}
\definecolor{codekw}{rgb}{0.85, 0.18, 0.50}
\definecolor{code}{RGB}{60, 80, 160}

The pseudocode of the manifold projector is shown in Algorithm~\ref{alg:manifold_projector}.

\begin{figure}[!th]
\begin{minipage}{\textwidth}
\begin{algorithm}[H]
\caption{Manifold Projector $\Psi$}
\label{alg:manifold_projector}
\begin{algorithmic}
\STATE \textcolor{codetext}{{\bfseries Input:} Dataset $\mathcal{Z}$, projector $\Psi$ with parameters $\theta_\Psi$, noise scales $\sigma_{1\ldots S}$, epochs $E$, learning rate $\eta$}
\FOR{$s=1$ to $S$}
    \STATE Initialize AdamW optimizer state $q$ and validation early stopping for this stage.
    \FOR{epoch$=1$ to $E$ or validation early stopping}
        \FOR{each minibatch $\mathcal{B}\subset\mathcal{Z}$}
            \FOR{each $z_i\in\mathcal{B}$}
                \STATE \textcolor{code}{$\epsilon_{i,1},\epsilon_{i,2}\overset{\mathrm{iid}}{\sim}\mathcal{N}(0,I)$}
                \STATE \textcolor{code}{$\tilde z_i^{(1)}\gets z_i+\sigma_s\epsilon_{i,1},\quad \tilde z_i^{(2)}\gets\tilde z_i^{(1)}+\sigma_s\epsilon_{i,2}$}
                \STATE \textcolor{code}{$\mathcal{L}_i\gets\|\Psi(\tilde z_i^{(1)})-z_i\|_2^2+\|\Psi(\tilde z_i^{(2)})-\tilde z_i^{(1)}\|_2^2$}
            \ENDFOR
            \STATE \textcolor{code}{$g\gets\nabla_{\theta_\Psi}\left(\frac{1}{|\mathcal{B}|}\sum_{i\in\mathcal{B}}\mathcal{L}_i\right)$}
            \STATE \textcolor{code}{$(\theta_\Psi,q)\gets\operatorname{AdamWUpdate}(\theta_\Psi,q,g;\eta)$}
        \ENDFOR
    \ENDFOR
\ENDFOR
\end{algorithmic}
\end{algorithm}
\end{minipage}
\end{figure}

\vspace{12pt}
\section{Hyperparameters and Architecture}
\label{supp:hyperparameters}

\paragraph{Training the organized autoencoder~(OAE)}
We trained the OAE using the AdamW optimizer with an initial learning rate of $10^{-3}$, together with a linear warmup cosine annealing scheduler. The learning rate was linearly increased from $10^{-5}$ to the target value during the first $10\%$ of training epochs~(warmup), and then cosine-annealed to $10^{-4}$ over the remaining epochs.  We used a batch size of $128$, a maximum of $100$ epochs, and early stopping with a patience of $20$ epochs.

\paragraph{Training the manifold projector}
The manifold projector is an MLP with widths $128\!\to\!32\!\to\!16\!\to\!32\!\to\!128$, ReLU hidden activations, and a linear output. We trained it using AdamW with a learning rate of $10^{-3}$. We used a batch size of $128$, a maximum of $100$ epochs per noise stage, and applied early stopping with a patience of $20$ epochs based on the validation loss. The same projector was trained successively at noise scales $0.5$, $0.2$, and $0.1$, with two independent Gaussian corruptions at the current scale for each clean embedding.

\textbf{Learning the manifold with SUGAR}
Prior to training the manifold projector $\Psi$, we hole-filled the OAE-encoded training latent space with SUGAR. To estimate degrees~(and thus sparsity), we used the $k$-NN mode with an $\alpha$-decay kernel ($\alpha{=}2$) and an adaptive bandwidth given by the distance to the $5$ nearest neighbors. The diffusion time was set to $t{=}1$. This $k$-NN degree estimator is distinct from the $k$-NN manifold retractor considered in our projector ablation study.

\begin{table}[h!]
\caption{Hyperparameters used for different datasets.
All runs use latent dimension $128$, denoising autoencoder manifold projector with $\mathrm{sugar}_w{=}1$, and $100$ Langevin steps.}
\centering
\resizebox{\textwidth}{!}{%
    \begin{tabular}{lccc}
    \toprule
    \textbf{Dataset} & \boldmath$\lambda_{\text{Recon}}$ / \boldmath$\lambda_{\text{Pred}}$ & \textbf{Noise levels} & \textbf{Langevin~(step size, temperature)} \\
    \midrule
    \texttt{OpenVaccine}     & 5.0 / 1.0 & \{0.5, 0.2, 0.1\} & $1\times 10^{-2},\, 1\times 10^{-3}$ \\
    \texttt{Zebrafish}       & 5.0 / 1.0 & \{0.5, 0.2, 0.1\} & $1\times 10^{-3},\, 1\times 10^{-3}$ \\
    \texttt{Ribosome}        & 5.0 / 1.0 & \{0.5, 0.2, 0.1\} & $1\times 10^{-3},\, 1\times 10^{-3}$ \\
    \bottomrule
    \end{tabular}
    }
\end{table}

\paragraph{Organized Autoencoder~(OAE)}
The OAE maps mRNA sequences $x\in\mathbb{R}^{L\times V}$ to a compact continuous latent $z\in\mathbb{R}^{d}$ with $d{=}128$ on all datasets.

The encoder uses a \texttt{Conv1d} stem followed by two downsampling stages of residual 1D blocks with \texttt{GroupNorm}, \texttt{SiLU}, and channel squeeze-excitation, doubling channels at each stride-2 stage. Features are pooled with \texttt{AdaptiveAvgPool1d} to $16$ positional bins and linearly projected to $z$. The property head is a three-layer MLP with \texttt{SiLU} and dropout $0.1$.

Decoding expands $z$ to a seed feature map, then applies a stack of \texttt{upsample} blocks (linear $\times 2$ + residual \texttt{Conv1d}) until length $\ge L$, interpolates to exact $L$ if needed, and refines with two residual \texttt{Conv1d} blocks to produce token logits. Weights use Kaiming/Xavier initialization. \texttt{GroupNorm} scales are set to $1$ and biases to $0$.

\paragraph{mRNA sequence vocabulary}
Although mRNA sequences naturally consist of the nucleotides \texttt{A}, \texttt{U}, \texttt{G}, and \texttt{C}, some experimental datasets represent \texttt{U} as \texttt{T}~(borrowing the DNA alphabet). To handle this heterogeneity consistently, we define a unified vocabulary of size $V=7$: \texttt{<pad>}, \texttt{A}, \texttt{G}, \texttt{C}, \texttt{T}, \texttt{U}, and \texttt{N}. Here, \texttt{N} denotes an unknown nucleotide during sequencing, and both \texttt{U} and \texttt{T} tokens are retained to ensure compatibility across datasets. Note that each dataset uniquely uses either \texttt{U} or \texttt{T}, and since we are training the model separately on each dataset, we do not risk confusing the model with interchangeability during decoding.

\paragraph{Hardware}
The evaluations were performed on a single NVIDIA A100 GPU. Generation timing covers optimization and decoding; model loading and oracle evaluation are excluded. For \ours, seed encoding is performed before timing.

\paragraph{Baseline adaptation}
MFM~\cite{MFM} was originally designed for smooth interpolation between populations on a manifold rather than direct property optimization. To adapt MFM to our optimization setting, we partition the training data into property tertiles, as required by the MFM formulation. We then train MFM to model flows from the low-property subpopulation to the high-property subpopulation for property maximization, and in the reverse direction for property minimization. The trained MFM model is subsequently applied to the unseen test set for property optimization, following the same evaluation protocol used for the other methods.

For PCD~\cite{PCD}, we use a multinomial restricted Boltzmann machine with categorical nucleotide variables and a property predictor trained on sequence labels. During optimization, property gradients with respect to one-hot nucleotide representations bias the visible-state logits in Gibbs sampling initialized from the input sequence. For EM~\cite{EM}, we train a sequence autoencoder and learn the energy potential on standardized latent embeddings. Optimization starts from the encoded input sequence and uses Langevin updates guided by a property-target loss evaluated through the decoder and a sequence property predictor. For MPGD~\cite{MPGD}, we use a sequence autoencoder and a latent diffusion model. We partially noise the encoded input sequence and guide the clean latent estimates during reverse diffusion using a property-target loss through the decoder. These adaptations retain the respective sampling mechanisms while enabling property-directed RNA sequence optimization.

DiffAb~\cite{DiffAb}, IgLM~\cite{IgLM}, NOS-C~\cite{NOS}, NOS-D~\cite{NOS}, and gg-dWJS~\cite{gg_dWJS} are implemented with the RNA vocabulary and trained on each dataset as described in the main text.

\section{Evaluation Metrics}
\label{supp:eval_metrics}

\paragraph{Dataset partitioning}

For the \texttt{Ribosome} dataset, the original paper provides predefined training and test splits. We use the provided test split as the test set and further partition the predefined training split into training and validation sets using an 85\%:15\% ratio. For the \texttt{OpenVaccine} and \texttt{Zebrafish} datasets, which do not come with predefined splits, we randomly partition the data into train, validation, and test sets using a 65\%:15\%:20\% ratio.

\paragraph{Property Optimization}
To quantify the effectiveness of property optimization of different models, we measure both the median property gain and the fraction of sequences that are successfully optimized.

Specifically, given a test set of mRNA sequences $X_\text{test}$ with predicted properties $\mathcal{P_\text{oracle}}(X_\text{test})$ and their optimized counterparts $\tilde{X}_\text{test}$ with properties $\mathcal{P}_\text{oracle}(\tilde{X}_\text{test})$, we compute:
\begin{equation}
\Delta_\text{median} = \operatorname{median}\big(\mathcal{P}_\text{oracle}(\tilde{x}) - \mathcal{P}_\text{oracle}(x)\big), \quad x \in X_\text{test}
\end{equation}
and
\begin{equation}
\%_\text{success} = \frac{1}{|X_\text{test}|} \sum_{x \in X_\text{test}} \mathds{1} \big[ \mathcal{P}_\text{oracle}(\tilde{x}) > \mathcal{P}_\text{oracle}(x) \big] \times 100\%
\end{equation}

Here, $\Delta_\text{median}$ measures the gain in the oracle-predicted target property across the test set, while $\%_\text{success}$ reports the percentage of sequences that improve after optimization.

For models that cannot refine existing sequences~(e.g., pure \textit{de novo} generators), we assign a random pairing between initial and final sequences to enable a fair comparison.

\begin{table}[!th]
\caption{The UTR-LM oracle for property evaluation.}
\centering
\resizebox{0.8\textwidth}{!}{%
    \begin{tabular}{cccc}
    \toprule
    & \multicolumn{2}{c}{Property Prediction} \\
    \cmidrule(lr){2-3}
    & Pearson Corr.~$\uparrow$ & Spearman Corr.~$\uparrow$ \\
    \midrule
    \texttt{OpenVaccine~(n$\approx$2k)} & 0.62 & 0.73 \\
    \texttt{Zebrafish~(n$\approx$55k)} & 0.82 & 0.81 \\
    \texttt{Ribosome~(n$\approx$260k)} & 0.93 & 0.91 \\
    \bottomrule
    \end{tabular}
}
\label{tab:oracle}
\end{table}

\paragraph{The external oracle}
In our evaluation for property optimization, we use an external oracle $\mathcal{P}_{\text{oracle}}(x)$ to predict the property of sequences generated by \ours as well as competing methods. Specifically, we adopt UTR-LM~\cite{UTR_LM}, a widely used mRNA language model, as the property prediction oracle. It is first pre-trained on more than 480,000 5' UTR mRNA sequences from diverse sources, and then fine-tuned for property prediction on each of our datasets. After fine-tuning, it is frozen and never accessed during optimization, ensuring unbiased evaluation. Performance of the fine-tuned UTR-LM is shown in Table~\ref{tab:oracle}.

For the third-party checkpoints used to produce the results in Table~\ref{tab:supp_main_different_oracle}, UTR-LM-TE is available at \url{https://huggingface.co/multimolecule/utrlm-te_el} and UTR-LM-MRL is available at \url{https://huggingface.co/multimolecule/utrlm-mrl}.

\section{Property Gradient Analysis}
\label{supp:property_gradient}

A central motivation of \ours is that property optimization along the data manifold should provide more reliable guidance than in the unconstrained ambient latent space. To examine predictor behavior in relation to training-data support, we compare gradient norms of the OAE property head and prediction variance among nearest training-set neighbors across three regimes in the OAE latent space:
\begin{itemize}[leftmargin=16pt, topsep=0pt, itemsep=4pt, parsep=0pt]
    \item \textbf{Densely sampled on-manifold regions:} training-set latent representations whose local density falls within the top $30\%$, estimated using adaptive $k$NN graph degree.
    \item \textbf{Sparsely sampled on-manifold regions:} training-set latent representations whose local density falls within the bottom $30\%$.
    \item \textbf{Off-manifold regions:} we perturb training-set latent vectors with isotropic Gaussian noise of scale $5\times$ the median on-manifold nearest-neighbor distance, then retain only those candidates whose distance to the nearest training latent exceeds $2.5\times$ the $95$th percentile of on-manifold nearest-neighbor distances~(falling back to the farthest candidates if fewer than $N$ points pass). This procedure constructs perturbed points far from the training embeddings.
\end{itemize}

Concretely, we encode the training set, subsample a reference dataset of up to $10{,}000$ latent vectors, and draw $N{=}500$ points per regime.
For each point we report local density, distance to the nearest training latent, the raw $\ell_2$ norm $\|\nabla_z f(z)\|_2$ of the property head, and the variance of $f$ over the $10$ nearest neighbors in the reference dataset.

Table~\ref{tab:property_gradient} summarizes predictor behavior across the three regions. For each seed, we compute the mean and standard deviation of each metric across the 500 points in each region, then average these means and standard deviations separately over five seeds. Thus, the reported $\pm$ values primarily describe within-region variability rather than variability across seeds.

Off-manifold points have lower mean gradient norms and greater variation in predictions among their nearest training-set neighbors. These observations characterize predictor behavior and neighborhood composition away from training-data support.

\vspace{16pt}
\begin{table}[!hb]
\caption{Property-gradient analysis in the OAE latent space on the Ribosome dataset. Reported values are the within-region mean and standard deviation averaged over five random seeds. Gradient norms are the raw $\ell_2$ norms $\|\nabla_z f(z)\|_2$ of the OAE property head.}
\label{tab:property_gradient}
\centering
\resizebox{\textwidth}{!}{%
    \begin{tabular}{lccccc}
    \toprule
    region & $N$ & density & nearest-neighbor distance & grad norm & prediction variance \\
    \midrule
    on-manifold, dense & 500 & $39.1\pm8.6$ & $6.273\pm0.428$ & $0.398\pm0.146$ & $0.127\pm0.103$ \\
    on-manifold, sparse & 500 & $\phantom{0}9.1\pm3.2$ & $8.254\pm0.669$ & $0.345\pm0.119$ & $0.182\pm0.142$ \\
    off-manifold & 500 & $\phantom{0}0.0\pm0.0$ & $399.3\pm25.7$ & $0.192\pm0.130$ & $0.583\pm0.430$ \\
    \bottomrule
    \end{tabular}
}
\end{table}

\clearpage
\newpage
\section{Additional Ablation Studies}
\label{supp:additional_ablation}

\begin{table}[!hb]
\caption{Ablation study on the design of the manifold projector $\Psi$. A simple $k$NN retractor is competitive in some settings but does not consistently support optimization in both directions. With a denoising autoencoder (DAE) as $\Psi$, multi-step denoising outperforms single-step denoising on \texttt{Zebrafish} and \texttt{Ribosome-loading}, and stays reasonably close on \texttt{OpenVaccine}.}
\label{supp:ablation_manifold_projector_design}
\centering
\resizebox{\textwidth}{!}{%
    \begin{tabular}{lcccccccccccc}
    \toprule
    & \multicolumn{4}{c}{\texttt{OpenVaccine (n$\approx$2k)}} & \multicolumn{4}{c}{\texttt{Zebrafish (n$\approx$55k)}} & \multicolumn{4}{c}{\texttt{Ribosome-loading (n$\approx$260k)}} \\
    \cmidrule(lr){2-5} \cmidrule(lr){6-9} \cmidrule(lr){10-13}
    & \multicolumn{2}{c}{$+$property} & \multicolumn{2}{c}{$-$property}
    & \multicolumn{2}{c}{$+$property} & \multicolumn{2}{c}{$-$property}
    & \multicolumn{2}{c}{$+$property} & \multicolumn{2}{c}{$-$property} \\
    \cmidrule(lr){2-3} \cmidrule(lr){4-5} \cmidrule(lr){6-7} \cmidrule(lr){8-9} \cmidrule(lr){10-11} \cmidrule(lr){12-13}
    $\Psi$ design
    & $\Delta$ $\uparrow$ & \% $\uparrow$ & $\Delta$ $\downarrow$ & \% $\uparrow$
    & $\Delta$ $\uparrow$ & \% $\uparrow$ & $\Delta$ $\downarrow$ & \% $\uparrow$
    & $\Delta$ $\uparrow$ & \% $\uparrow$ & $\Delta$ $\downarrow$ & \% $\uparrow$ \\
    \midrule
    $k$NN ($k{=}1$) & $+0.01$ & $53.2$ & $+0.00$ & $50.3$ & $-0.22$ & $45.8$ & $-0.42$ & $61.5$ & $+0.05$ & $55.5$ & $-0.09$ & $56.7$ \\
    $k$NN ($k{=}5$) & $+0.06$ & $56.2$ & $+0.05$ & $45.4$ & $-0.22$ & $46.0$ & $-0.37$ & $59.7$ & $-1.05$ & $\phantom{0}9.7$ & $-1.17$ & $92.4$ \\
    $k$NN ($k{=}10$) & $+0.10$ & $58.6$ & $+0.07$ & $43.1$ & $-0.22$ & $45.7$ & $-0.29$ & $58.2$ & $-1.09$ & $\phantom{0}8.9$ & $-1.21$ & $92.5$ \\
    DAE, single-step & $+\textbf{0.93}$ & $\textbf{88.3}$ & $-\textbf{0.79}$ & $\textbf{87.4}$ & $+0.24$ & $58.1$ & $-0.83$ & $72.1$ & $+0.75$ & $84.4$ & $-0.99$ & $74.4$ \\
    \rowcolor{YaleBlue!12} DAE, multi-step & $+0.92$ & $83.6$ & $-0.77$ & $84.2$ & $+\textbf{0.79}$ & $\textbf{70.9}$ & $-\textbf{1.25}$ & $\textbf{77.3}$ & $+\textbf{0.93}$ & $\textbf{88.8}$ & $-\textbf{1.93}$ & $\textbf{92.8}$ \\
    \bottomrule
    \end{tabular}
}
\end{table}

\begin{table}[!hb]
\caption{Ablation study on the reconstruction weight $\lambda_\text{Recon}$. Although $\lambda_\text{Recon}=1.0$ outperforms $\lambda_\text{Recon}=5.0$ on \texttt{OpenVaccine} in both optimization directions, we decide to use $\lambda_\text{Recon}=5.0$ across all datasets to maintain a consistent hyperparameter setting.}
\label{supp:ablation_recon}
\centering
\resizebox{\textwidth}{!}{%
    \begin{tabular}{lcccccccccccc}
    \toprule
    & \multicolumn{4}{c}{\texttt{OpenVaccine (n$\approx$2k)}} & \multicolumn{4}{c}{\texttt{Zebrafish (n$\approx$55k)}} & \multicolumn{4}{c}{\texttt{Ribosome-loading (n$\approx$260k)}} \\
    \cmidrule(lr){2-5} \cmidrule(lr){6-9} \cmidrule(lr){10-13}
    & \multicolumn{2}{c}{$+$property} & \multicolumn{2}{c}{$-$property}
    & \multicolumn{2}{c}{$+$property} & \multicolumn{2}{c}{$-$property}
    & \multicolumn{2}{c}{$+$property} & \multicolumn{2}{c}{$-$property} \\
    \cmidrule(lr){2-3} \cmidrule(lr){4-5} \cmidrule(lr){6-7} \cmidrule(lr){8-9} \cmidrule(lr){10-11} \cmidrule(lr){12-13}
    $\lambda_\text{Recon}$
    & $\Delta$ $\uparrow$ & \% $\uparrow$ & $\Delta$ $\downarrow$ & \% $\uparrow$
    & $\Delta$ $\uparrow$ & \% $\uparrow$ & $\Delta$ $\downarrow$ & \% $\uparrow$
    & $\Delta$ $\uparrow$ & \% $\uparrow$ & $\Delta$ $\downarrow$ & \% $\uparrow$ \\
    \midrule
    \phantom{0}0.0 & $-1.29$ & $16.2$ & $-2.40$ & $91.0$ & $-0.79$ & $33.9$ & $-0.34$ & $54.9$ & $+0.05$ & $54.0$ & $-0.21$ & $57.3$ \\
    \phantom{0}0.1 & $+1.00$ & $66.8$ & $-1.60$ & $88.5$ & $-0.11$ & $51.3$ & $-1.08$ & $76.1$ & $+0.39$ & $77.5$ & $-1.20$ & $90.6$ \\
    \phantom{0}0.5 & $+0.93$ & $87.2$ & $-\textbf{2.92}$ & $\textbf{99.4}$ & $+0.51$ & $63.7$ & $-1.07$ & $76.7$ & $+0.85$ & $88.7$ & $-0.98$ & $69.5$ \\
    \phantom{0}1.0 & $+\textbf{1.11}$ & $\textbf{91.4}$ & $-1.31$ & $87.8$ & $+\textbf{1.13}$ & $\textbf{81.2}$ & $-1.09$ & $77.2$ & $+0.71$ & $85.4$ & $-0.88$ & $69.6$ \\
    \rowcolor{YaleBlue!12} \phantom{0}5.0 & $+0.92$ & $83.6$ & $-0.77$ & $84.2$ & $+0.79$ & $70.9$ & $-\textbf{1.25}$ & $\textbf{77.3}$ & $+\textbf{0.93}$ & $\textbf{88.8}$ & $-\textbf{1.93}$ & $\textbf{92.8}$ \\
    10.0 & $-0.49$ & $49.2$ & $-0.57$ & $70.7$ & $+0.58$ & $66.4$ & $-0.98$ & $76.4$ & $+0.68$ & $82.0$ & $-0.06$ & $40.3$ \\
    \bottomrule
    \end{tabular}
}
\end{table}

\clearpage
\newpage
\section{Additional comparison against de novo baselines}
\label{supp:comparison_de_novo}

In addition to state-of-the-art biological sequence optimization methods, we compare \ours with two different types of methods. The first type is \textit{de novo} generative modeling approaches, namely variational autoencoder~(VAE)~\cite{VAE}, denoising diffusion probabilistic model~(DDPM)~\cite{DDPM}, latent diffusion model~(LDM)~\cite{LDM}, and flow matching~(FM)~\cite{FM}. The second type comprises classic optimization algorithms that operate in our OAE latent manifold, including gradient ascent~\cite{GradientAscent, zinkevich2003online}, Markov chain Monte Carlo~(MCMC)~\cite{brooks1998markov, MCMC_ML}, and hill climbing~\cite{selman2006hill}. All classic optimization baselines are our GPU-compatible re-implementations of Castro et al.~\cite{ReLSO}. Hill climbing moves toward the best-scoring neighboring training embedding; its stochastic variant samples an improving neighbor uniformly, or any neighbor if none improves the score.

\begin{table*}[!th]
\caption{\ours achieves better property optimization than \textit{de novo} generative methods and classic optimization algorithms that operate on our learned data manifold.}
\vspace{4pt}
\centering
\resizebox{0.7\textwidth}{!}{%
    \begin{tabular}{lcccc}
    \toprule
    \multirow{3}{*}{Method} & \multicolumn{4}{c}{Property Optimization} \\
    \cmidrule(lr){2-5}
    & \multicolumn{2}{c}{$+$Property} & \multicolumn{2}{c}{$-$Property} \\
    \cmidrule(lr){2-3} \cmidrule(lr){4-5}
    & $\Delta$~$\uparrow$ & \% $\uparrow$ & $\Delta$~$\downarrow$ & \% $\uparrow$ \\
    \midrule
    \multicolumn{5}{l}{\texttt{OpenVaccine~(n$\approx$2k)}} \\
    \cmidrule(lr){1-5}
    VAE~\cite{VAE} & $+0.23$ & $56.5$ & $+0.23$ & $43.5$ \\
    DDPM~\cite{DDPM} & $-0.33$ & $35.1$ & $-0.33$ & $64.9$ \\
    LDM~\cite{LDM} & $-0.41$ & $32.4$ & $-0.41$ & $67.6$ \\
    FM~\cite{FM} & $-0.28$ & $37.9$ & $-0.28$ & $62.1$ \\
    OAE + gradient ascent & $+0.00$ & $38.9$ & $+0.00$ & $40.0$ \\
    OAE + MCMC & $+0.02$ & $52.9$ & $+0.00$ & $48.9$ \\
    OAE + hill climbing & $+0.05$ & $57.5$ & $-0.10$ & $65.6$ \\
    OAE + stochastic hill climbing & $+0.02$ & $52.4$ & $-0.03$ & $55.3$ \\
    \rowcolor{YaleBlue!12} \textbf{\ours~(ours)} & $+\textbf{0.92}$ & $\textbf{83.6}$ & $-\textbf{0.77}$ & $\textbf{84.2}$ \\
    \midrule
    \multicolumn{5}{l}{\texttt{Zebrafish~(n$\approx$55k)}} \\
    \cmidrule(lr){1-5}
    VAE~\cite{VAE} & $-0.17$ & $46.0$ & $-0.17$ & $54.0$ \\
    DDPM~\cite{DDPM} & $+0.27$ & $57.5$ & $+0.27$ & $42.5$ \\
    LDM~\cite{LDM} & $-0.67$ & $36.0$ & $-0.67$ & $64.0$ \\
    FM~\cite{FM} & $+0.22$ & $56.7$ & $+0.22$ & $43.3$ \\
    OAE + gradient ascent & $-0.01$ & $48.6$ & $-0.02$ & $52.5$ \\
    OAE + MCMC & $-0.18$ & $43.9$ & $-0.25$ & $59.3$ \\
    OAE + hill climbing & $+0.17$ & $56.9$ & $-0.16$ & $56.4$ \\
    OAE + stochastic hill climbing & $+0.09$ & $54.5$ & $-0.06$ & $53.5$ \\
    \rowcolor{YaleBlue!12} \textbf{\ours~(ours)} & $+\textbf{0.79}$ & $\textbf{70.9}$ & $-\textbf{1.25}$ & $\textbf{77.3}$ \\
    \midrule
    \multicolumn{5}{l}{\texttt{Ribosome~(n$\approx$260k)}} \\
    \cmidrule(lr){1-5}
    VAE~\cite{VAE} & $-0.28$ & $43.0$ & $-0.28$ & $57.0$ \\
    DDPM~\cite{DDPM} & $-0.10$ & $45.8$ & $-0.10$ & $54.2$ \\
    LDM~\cite{LDM} & $-0.18$ & $44.6$ & $-0.18$ & $55.4$ \\
    FM~\cite{FM} & $-0.11$ & $45.6$ & $-0.11$ & $54.4$ \\
    OAE + gradient ascent & $+0.10$ & $57.0$ & $-0.04$ & $52.1$ \\
    OAE + MCMC & $+0.01$ & $50.6$ & $-0.01$ & $50.3$ \\
    OAE + hill climbing & $+0.07$ & $53.9$ & $-0.12$ & $60.2$ \\
    OAE + stochastic hill climbing & $+0.05$ & $53.7$ & $-0.09$ & $55.6$ \\
    \rowcolor{YaleBlue!12} \textbf{\ours~(ours)} & $+\textbf{0.93}$ & $\textbf{88.8}$ & $-\textbf{1.93}$ & $\textbf{92.8}$ \\
    \bottomrule
    \end{tabular}
}
\label{tab:ablation_optimization}
\end{table*}

\clearpage
\newpage
\section{Interpolation between sequences}
\label{sec:method_interpolate}

In addition to optimizing mRNA sequences, we can interpolate between two existing sequences by guiding the latent embedding of one sequence toward that of another. Specifically, given a source sequence $x_\text{source}$ and a target sequence $x_\text{target}$, we first obtain their latent embeddings via the encoder: $z_\text{source} = \mathcal{E}(x_\text{source})$ and $z_\text{target} = \mathcal{E}(x_\text{target})$.

We then perform property-guided Manifold Langevin dynamics starting from $z = z_\text{source}$, with the drift term being the force term that drives the embedding toward $z_\text{target}$:
\begin{equation}
f_{\text{interp}}(z, z_\text{target}) = -\frac{z - z_\text{target}}{\| z - z_\text{target} \|_2}
\label{eqn:normed_force}
\end{equation}
In this case, we slightly modify the update rule of the Langevin dynamics~(Eqn~\eqref{eqn:update_rule}) as follows:
\begin{equation}
\label{eqn:langevin_interpolate}
\text{d}z_{t} = \eta \left(\frac{1}{\tau}\nabla_z f(z_t) + \lambda f_{\text{interp}}(z_t,z_\text{target}) \right) + \sqrt{2\eta} \cdot \epsilon_t.
\end{equation}
By setting the interpolation weight $\lambda > 0$ in Eqn~\eqref{eqn:langevin_interpolate}, we add a directional bias that drives the latent trajectory toward the target point, while following the property gradients.

\begin{figure*}[!ht]
\centering
\includegraphics[width=\textwidth]{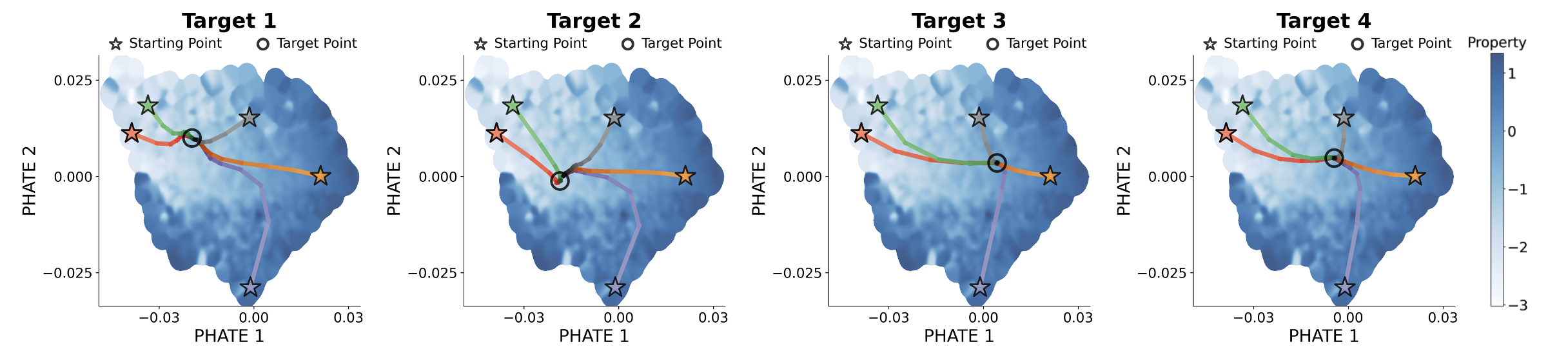}
\caption{Latent space interpolation trajectories from 5 sources to 4 targets. Each trajectory is shown as a line fading from bright to dark in a consistent color. \ours produces smooth and coherent paths on the manifold between arbitrary input-target mRNA pairs.}
\label{fig:latent_space_interpolation}
\end{figure*}

\begin{figure}[!ht]
\centering
\includegraphics[width=\textwidth]{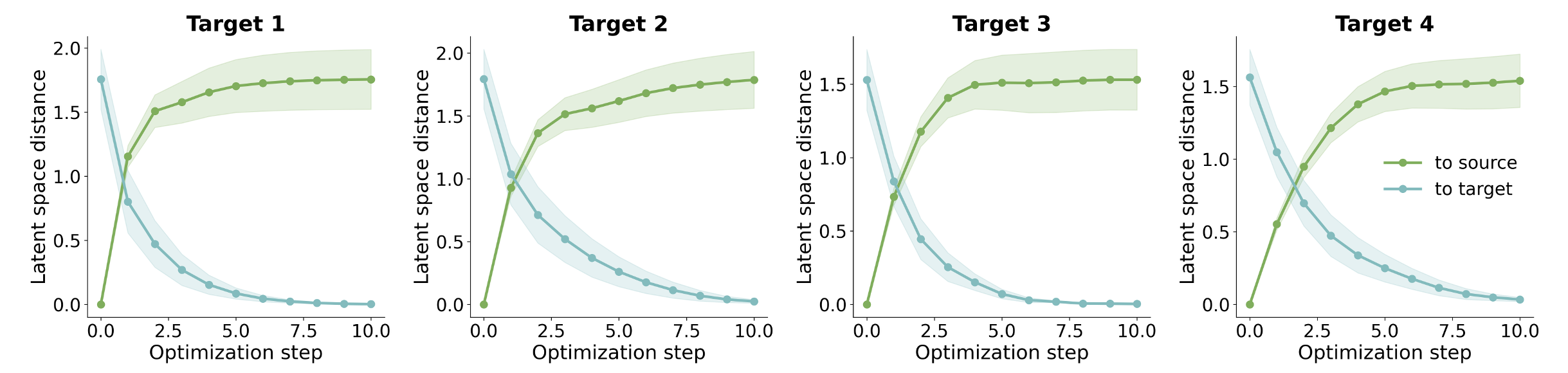}
\caption{Latent space $\ell_2$ distances during interpolation show smooth and monotonic transition from the source to the target. Results are averaged over all data samples.}
\label{fig:interpolation_distance}
\end{figure}

\paragraph{Results on interpolation}
Guided by a directional force toward a specified target, \ours generates smooth and coherent trajectories on the learned manifold while preserving biological plausibility and continuity~(Figure~\ref{fig:latent_space_interpolation}). These trajectories connect arbitrary input-target sequence pairs in a structured manner, reflecting semantically meaningful transitions. The distances from each intermediate point to the source and target quantitatively demonstrate the monotonicity and smoothness of the interpolation~(Figure~\ref{fig:interpolation_distance}).

\clearpage
\newpage
\section{Additional optimization trajectories}
\label{supp:additional_trajectories}

\begin{figure*}[!th]
\centering
\includegraphics[width=\textwidth]{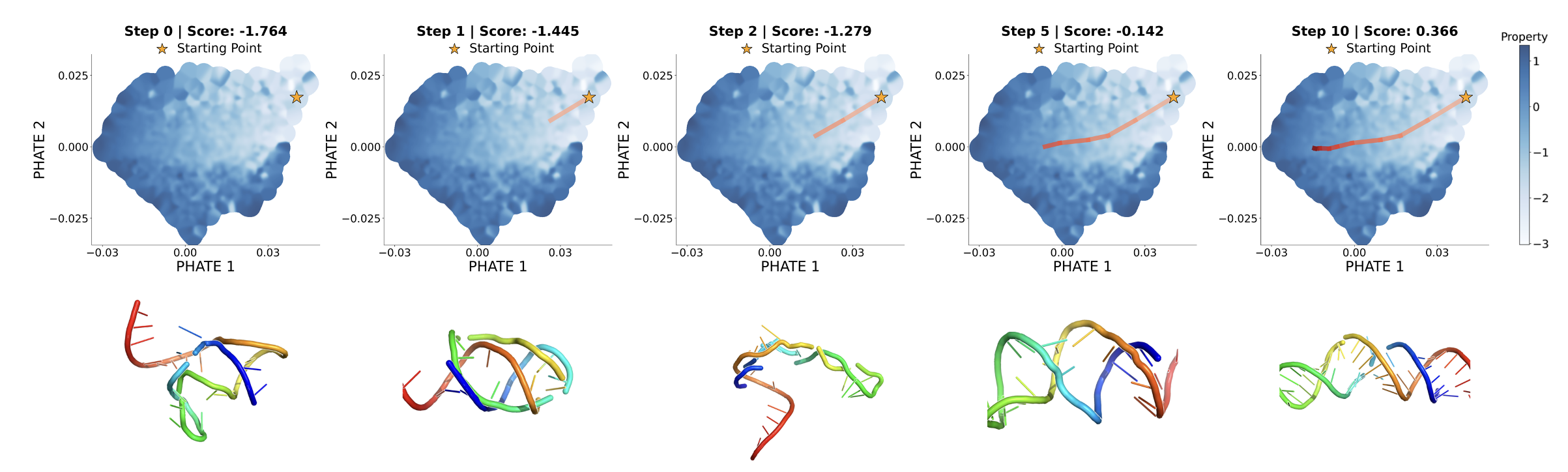}
\caption{More examples of latent manifold trajectories. Trajectories in PHATE space and corresponding 3D structures are shown.}
\label{fig:latent_space_trajs2}
\end{figure*}

\begin{figure*}[!th]
\centering
\includegraphics[width=\textwidth]{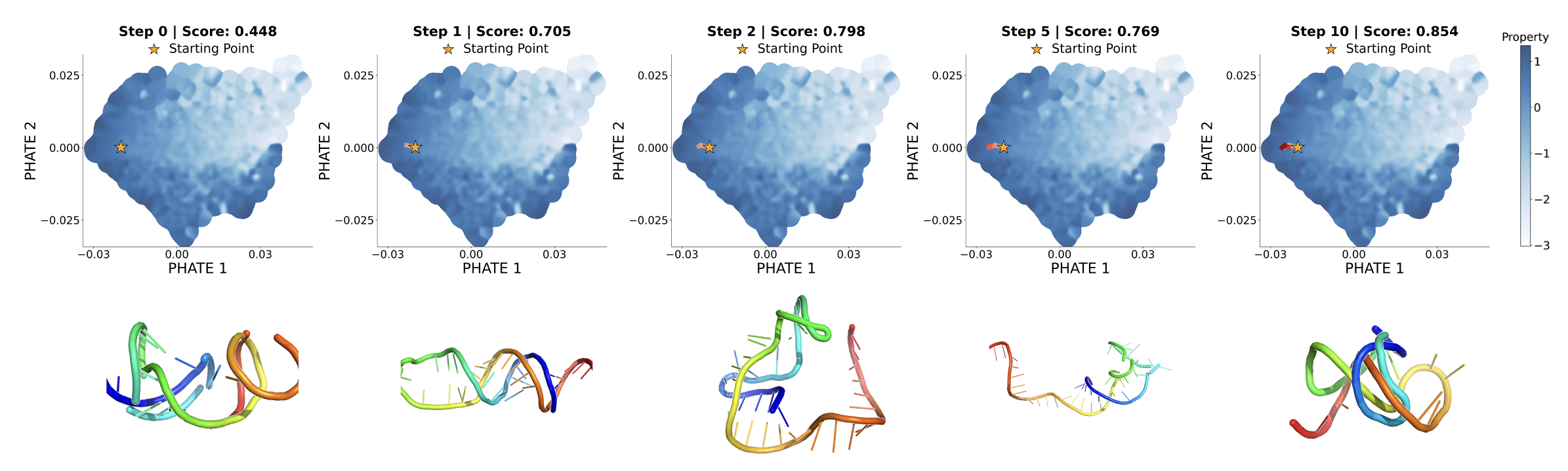}
\caption{More examples of latent manifold trajectories. Trajectories in PHATE space and corresponding 3D structures are shown.}
\label{fig:latent_space_trajs3}
\end{figure*}

\begin{figure*}[!th]
\centering
\includegraphics[width=\textwidth]{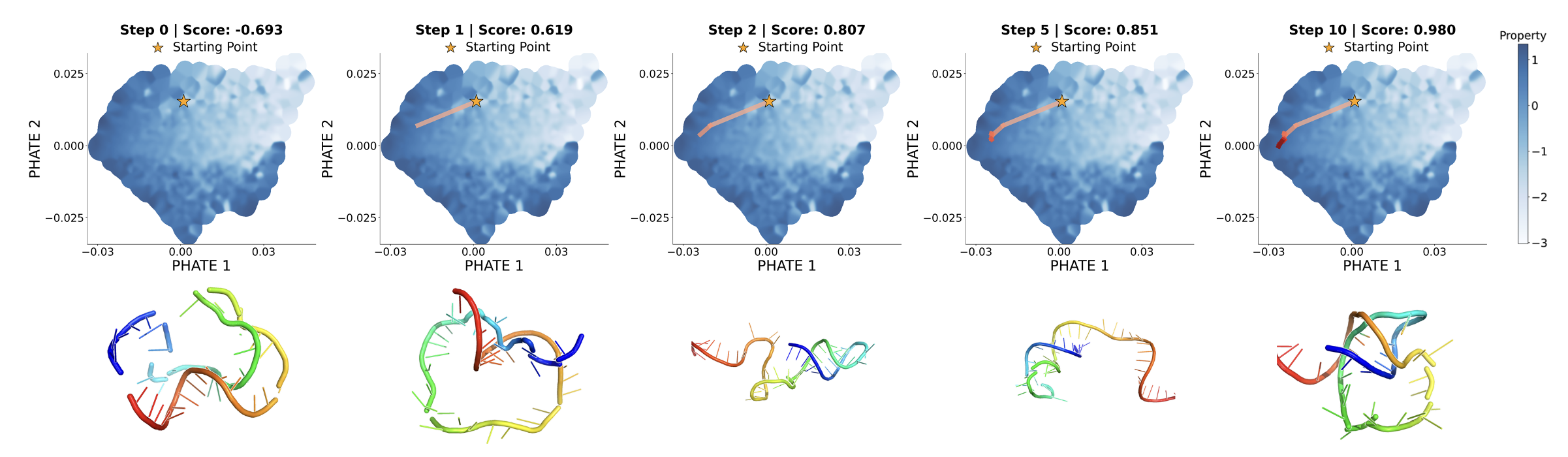}
\caption{More examples of latent manifold trajectories. Trajectories in PHATE space and corresponding 3D structures are shown.}
\label{fig:latent_space_trajs4}
\end{figure*}

\clearpage
\newpage
\section{Limitations and Future Work}
\label{supp:limitations}
One limitation of \ours is that it models and optimizes mRNA sequences only, without explicitly incorporating mRNA structure. While sequence-based representations capture many aspects of mRNA function, structures can provide auxiliary information in regulating translation, stability, and interactions.

In future work, we will study the possibility of performing sequence-structure joint modeling and optimization. Beyond mRNA, we plan to extend \ours to other modalities such as protein sequences and regulatory elements, and integrate active learning frameworks that guide wet lab experimentation. Such extensions would require a domain-appropriate encoder-decoder and training on domain-specific data, while retaining the overall formulation of property-organized latent learning, manifold projection, and property-guided Langevin optimization. By grounding sequence optimization on the manifold of real data, we aim to provide a versatile platform for interpretable high-throughput design in synthetic biology.

\section{Impact Statement}
\label{supp:impact}

This work proposes a machine learning framework for optimizing mRNA sequences while preserving biological viability proxies by constraining optimization to a learned data manifold. Potential positive impacts include supporting computational design workflows in mRNA therapeutics and synthetic biology such as vaccine development by enabling more controlled and interpretable sequence optimization. As a general machine learning contribution, the framework may also inform broader research on manifold-constrained optimization for biological sequences.

The approach is intended as a computational tool to complement experimental and domain expertise rather than a substitute for empirical validation. As with other data-driven sequence models, limitations and biases in the available training data may influence optimization outcomes, and inappropriate use without biological oversight could lead to nonfunctional or misleading designs.

We do not anticipate direct negative societal consequences arising from this work beyond those already associated with machine learning-based biological modeling. We encourage future research to further study robustness, data bias, and safe integration of such methods into experimental pipelines.

\end{appendices}


\end{document}